\documentclass[
reprint,
showpacs,
preprintnumbers,
nofootinbib,
amsmath,
amssymb,
aps,
prd,
]{revtex4-1}
\usepackage{enumerate}
\usepackage{color}
\usepackage{graphicx}
\usepackage{dcolumn}
\usepackage{bm}
\usepackage{empheq}
\usepackage{nomencl}
\usepackage{hyperref}
\usepackage{mathrsfs}
\usepackage{amsmath}
\hypersetup
{
	colorlinks,%
	citecolor=blue,%
	linkcolor=magenta,%
	urlcolor=blue,%
}

\usepackage{subfigure}

\begin{document}	
	\title{Lensing and light rings of parity-odd rotating boson stars}
	\author{Yang Huang $^1$}\email{sps\_huangy@ujn.edu.cn}
	\author{Dao-Jun Liu $^2$}\email{corresponding author djliu@shnu.edu.cn}
	\author{Hongsheng Zhang $^{1}$}\email{corresponding author sps\_zhanghs@ujn.edu.cn}
	\affiliation{
		$^1$ School of Physics and Technology, University of Jinan, Jinan 250022, China}
	\affiliation{
		$^2$ Department of Physics, Shanghai Normal University, Shanghai 200234, China}

	\begin{abstract}
	We obtain the first image of a parity-odd celestial body. Recently, an intriguing parity-odd rotating boson star was proposed. We investigate the lensing effects of these stars in detail. Our analysis demonstrates distinct gravitational distortions around these stars, clearly differentiating them from their parity-even counterparts. Furthermore, we analyze the conditions under which chaotic behavior appears in the images of ultra-compact stars.
	\end{abstract}
	\maketitle

	
	\section{Introduction}
	
	Traditionally, all the ingredients are treated as fluids in the studies of star structures. This approach is also applicable to the structures of objects in astrophysics, ranging from planets to galaxies, and even the whole Universe. The fluid approximation is an acceptable approach at  macroscopic and low energy level. However, fundamentally, we should study the stuffs in the universe based on field theory and general relativity. This more fundamental approach can be traced back to Wheeler's geon, where electromagnetic fields are self-gravitated to form a lumpy structure \cite{Wheeler:1955zz,Wheeler:1957}. Unfortunately, this structure is unstable against linear perturbations.
	
	In the late 1960s, it was realized that stable localized soliton-type configurations, now dubbed boson stars (BSs),  may arise when a complex scalar field is coupled to gravity \cite{Kaup1968,Ruffini:1969qy}. Since then, BSs have found applications in a wide variety of fundamental physics and astrophysics models, including sources of dark matter, black hole mimickers, sources of gravitational waves in  binary systems and other phenomena. For comprehensive reviews, see, e.g., Refs. \cite{Jetzer1992,Schunck2003,Visinelli2021,Liebling2023,Khlopov:1985fch}. The model in which a scalar is minimally coupled to gravity also takes remarkable status in investigations of gravitational collapse \cite{Guo:2018yyt,Guo:2020ked,Hu:2023qcq}.
	
	The potential of BSs to serve as halos of galaxies or clusters of galaxies, and to mimic binary black holes that excite gravitational waves, is being seriously investigated. BSs evade the intractable problem of singularity inherent in black holes while producing wave forms similar to those detected at gravitational wave observatories. As they can play the role of halos, they are also significant for dark matter research and new physics. The only scalar particle in the standard model is the Higgs boson, which is not stable and thus may not be an appropriate candidate for the scalar in studies of BSs. The studies of BSs almost inevitably lead to new physics, providing valuable hints for theories beyond the standard model.
	
	The properties of BSs are primarily determined by the potential that encodes the self-interaction of the complex scalar field and varies with the Lagrangian form \cite{Mielke1981,Colpi:1986ye,Lee1987,Li:2001he,Guerra2019}. The simplest BS solutions are the ones originally obtained \cite{Kaup1968,Ruffini:1969qy} and coined mini-boson star later \cite{Lee1989}. These solutions possess spherical symmetry and arise in the massive Einstein-Klein-Gordon theory with global U(1) symmetry, representing gravitationally bounded lumps of scalar condensate. Indeed, depending on the strength of the coupling  and the other parameters of the self-interaction potential, the size and mass of the BSs can vary from atomic to astrophysical scales. Many generalizations of the original BS solutions have been proposed, such as Proca stars, where scalar fields are replaced by massive vector fields \cite{Brito2016},  black holes with bosonic hair, where BSs define the boundary in the parameter space \cite{Herdeiro2015}, Fermion-boson stars \cite{Henriques1989}, multi-state BSs \cite{Bernal2010,Li:2019mlk,Li:2020ffy,Zeng:2023hvq}, axion stars \cite{Braaten:2015eeu,Zhang:2018slz,Braaten:2019knj}, and multi-field BSs \cite{Hawley2003}.
	
	Another kind of important generalization of BSs involves the introduction of rotation. The first solutions for rotating boson stars (RBS) in general relativity were obtained in the mid-1990s \cite{Schunck1996,Yoshida1997}. These solutions were derived under the assumption of an isotropic, stationary, axisymmetric metric of spacetime for a system in which a free scalar field is minimally coupled to Einstein’s gravity. Notably, they exhibited an important quantization relation $J= n Q$, where $J$ is the angular momentum, $Q$ is the scalar charge, and $n$ must be an integer.
	
	The field equations allow for BSs with both positive and negative parity, where the scalar field is symmetric and antisymmetric with respect to reflections, respetively \cite{Volkov2002}. BSs with positive (negative) parity are also referred to as parity-even (odd) BSs. Parity-odd RBSs were first studied by Kleihaus et al. \cite{Kleihaus:2007vk}. It was later discovered that a rotating black hole forms inside a parity-odd RBS as the BS becomes increasingly compact, leading in what is known as a parity-odd rotating hairy black hole \cite{Wang:2018xhw,Kunz:2019bhm}. One of the most intriguing features of parity-odd BSs is that they may exhibit ergo-double-tori, $\left(S^1\times S^1\right)\oplus\left(S^1\times S^1\right)$, which have significant implications for the geodesic structure of the surrounding spacetime.
	
	Although there exist a number of similarities between BSs and other compact objects such as black holes and neutron stars, BSs  are horizonless, transparent for photons and singularity-free, have no sharp stellar surface, and may be distinguished observationally from both black holes and neutron stars. Gravitational wave signatures of highly compact BS binaries have been investigated by Palenzuela et al. \cite{Palenzuela2017}. The study reveals that the overall gravitational wave signals from solitonic BSs are unlikely to degenerate with those from either binary neutron stars or binary black holes, except possibly in the most compact cases. It has been recently pointed out that the prolonged gravitational wave afterglow can provide a characteristic signal that may distinguish it from other astrophysical sources \cite{Croft2023}. The lensing effects (image) is a systematic and powerful method for observations of BSs. The possibility of BSs acting as gravitational lenses and their lensing effects have also been studied \cite{Dabrowski:1998ac,Rosa:2022toh,Rosa:2022tfv,Rosa:2023qcv,Rosa:2024eva}.
	
	It is found that the phenomenology associated with the light ring reveals remarkable chaotic lensing patterns \cite{Cunha:2016bjh}, which are quite distinct from a standard black hole shadow (see, e.g., Ref. \cite{Chen:2022scf} for a review). The strong lensing region can be considerably smaller than the shadow of a black hole with a comparable mass \cite{Cunha2017}. The light ring also plays an important role in the scattering process of gravitational waves \cite{Hongsheng:2018ibg}. With the continued advancement in observations, both in the electromagnetic and gravitational spectra, we may have evidence for the existence of BSs in the near future.
	
	While the observational signatures of parity-even BSs have been explored in-depth, those of parity-odd BSs receive less attention. It is a distinctive property that there exists parity-odd object. Recently, however, there has been increasing interest in parity-odd BSs. These are now interpreted as two spinning BSs in equilibrium, where the gravitational attraction between them is balanced by a repulsive scalar interaction \cite{Herdeiro:2023roz}. This repulsive interaction arises from a phase difference of $\pi$ between the northern and southern hemispheres. Interestingly, it has been shown that, in some cases, parity-odd rotating black holes may possess three equilibrium points along the rotation axis. At these points, spinning black holes can emerge and grow, resulting in either a single spinning hairy black hole \cite{Kunz:2019bhm} or two spinning hairy black holes \cite{Herdeiro:2023roz}, which are balanced by the surrounding scalar field.
	
	In this paper, we thoroughly investigate the lensing and light rings effects of these parity-odd rotating BSs. The rest part of the paper is organized as follows. First, in Sec. \ref{Sec: setup} we reconstruct the parity-odd BS solutions first presented in \cite{Kunz:2019bhm} and achieve high accuracy in reproducing their results. Next, in Sec. \ref{Sec: Lensing} and  we analyze the lensing effects and the light ring structure of the numerical solutions. Finally, we conclude this paper in Sec. \ref{Sec: conclude}.	
	
	\section{Odd-parity rotating boson stars}\label{Sec: setup}
	\subsection{The model}
	In this section, we reconstruct the numerical solutions of parity-odd RBS. We focus on the model in which a massive scalar field with mass $\mu$ is minimally coupled to gravity
	\begin{equation}
		S=\int d^4x\sqrt{-g}\left(\frac{R}{2\kappa}-\mathcal{L}_m\right),
	\end{equation}
	where $\kappa=8\pi$, $R$ is the Ricci scalar, and $\mathcal{L}$ is the Lagrangian
	\begin{equation}
		\mathcal{L}_m=g^{\alpha\beta}\Psi^*_{,\alpha}\Psi_{,\beta}+\mu^2\Psi^*\Psi.
	\end{equation}
	The resulting equations of motion are given by
	\begin{equation}\label{Eq: EOM}
		\begin{aligned}
			R_{\alpha\beta}-\frac{1}{2}g_{\alpha\beta}R&=\kappa T_{\alpha\beta},\\
			\left(\Box-\mu^2\right)\Psi&=0,
		\end{aligned}
	\end{equation}
	where
	\begin{equation}
		T_{\alpha\beta}=\Psi^*_{,\alpha}\Psi_{,\beta}+\Psi^*_{,\beta}\Psi_{,\alpha}-g_{\alpha\beta}\mathcal{L}_m
	\end{equation}
	is the stress-energy tensor of the scalar field. It can be shown that the system is invariant under the global $U(1)$ transformation of the scalar field, $\Psi\rightarrow\Psi e^{i\chi}$, where $\chi$ is a constant. As a result, the scalar $4$-current, $j^\alpha=-i\left(\Psi^*\partial^\alpha\Psi-\Psi\partial^\alpha\Psi^*\right)$, is conserved, and the corresponding conserved charge is given by
	\begin{equation}
		Q=\int d^3x\sqrt{-g}j^t.
	\end{equation}
	
	Rotating BSs are described by the metric \cite{Herdeiro:2014goa,Herdeiro:2015gia}
	\begin{equation}\label{Eq: ansatzMetric}
		\begin{aligned}
			ds^2=&-e^{2F_0}dt^2+e^{2F_1}\left(dr^2+r^2d\theta^2\right)\\
			&+e^{2F_2}r^2\sin^2\theta\left(d\varphi-Wdt\right)^2,
		\end{aligned}
	\end{equation}
	where $\{F_0, F_1, F_2, W\}$ are functions of $(r,\theta)$. The axially symmetric ansatz for the stationary scalar field is
	\begin{equation}\label{Eq: ansatzScalar}
		\Psi=\phi(r,\theta)e^{i(m\varphi-wt)},
	\end{equation}
	where $w$ is the scalar field frequency and $m=\pm1,\pm2,\cdots$ is the azimuthal harmonic index. Without loss of generality, one can assume $w>0$. In this paper, we focus on solutions with $m=1$ for simplicity.
	
	The partial differential equations for $\phi$, $F_i\ (i=0,1,2)$, and $W$ can be found in Eqs. (2.10) and (2.11) of \cite{Herdeiro:2015gia}, and we solve them using boundary conditions that are almost the same, except the scalar field on the equatorial plane. Since we deal with odd-parity solutions where
	\begin{equation}\label{Eq: BC1}
		\partial_\theta F_i=\partial_\theta W=\phi=0,\;\;\text{at},\;\;\theta=\pi/2,
	\end{equation}
	and
	\begin{equation}\label{Eq: BC2}
		\partial_\theta F_i=\partial_\theta W=\phi=0\;\;\text{at}\;\;\theta=0,\;\text{or}\;\pi.
	\end{equation}
	
	\subsection{Numerical approach}
	It is useful to introduce a new radial coordinate \cite{Kunz:2019bhm}
	\begin{equation}\label{Eq: copact coord}
		x=\frac{r}{r+L},
	\end{equation}
	which maps the semi-infinite region $[0,\infty)$ to a compact interval $[0,1]$. Here $L$ is an appropriate constant which is used to adjust the location of numerical grids in the radial direction. The system of equations resulting from Eq.(\ref{Eq: EOM}) is discretized on a uniform grid
	\begin{equation}
		\begin{aligned}
			x_i&=\frac{i}{N_x},\;\;i=0,1,\cdots,N_x,\\
			\theta_j&=\frac{(2j+1)\pi}{4N_\theta},\;\;j=0,1,\cdots,N_\theta-1.
		\end{aligned}
	\end{equation}
	In most cases, we have chosen $(N_x,N_\theta)=(100,8)$. In this framework, a fourth-order finite difference method is employed in the radial direction, while a spectral method based on Fourier series expansion is used in the angular direction. The metric functions $F_1, F_2, F_0, W$ and the scalar field $\phi$ are expanded using different Fourier basis functions based on their respective angular boundary conditions specified in equations (\ref{Eq: BC1}) and (\ref{Eq: BC2}) \cite{Fernandes:2022gde}. The expansions are given by:
	\begin{equation}\label{Eq: Expansion Met}
		u(x,\theta)=\frac{1}{2}\tilde{u}_0(x)+\sum^{N_\theta-1}_{n=1}\tilde{u}_n(x)\cos\left(2n\theta\right),
	\end{equation}
	where $u=\{F_1,F_2,F_0,W\}$ and
	\begin{equation}
		\phi(x,\theta)=\sum^{N_\theta-1}_{n=0}\tilde{\phi}_n(x)\sin(2n\theta).
	\end{equation}
	The coefficients $\tilde{u}$ in Eq.(\ref{Eq: Expansion Met}) are constructed from the numerical function values on the grid $u_{ij}\equiv u(x_i,\theta_j)$ as follows. First, we compute each coefficient using the formula,
	\begin{equation}
		\tilde{u}_n(x_i)=\frac{2}{N_\theta}\sum_{j=0}^{N_\theta-1}u_{ij}\cos(2n\theta_j).
	\end{equation}
	Then, we interpolate $\tilde{u}_{n}(x)$ using Newtonian interpolation method. This approach ensures the accuracy and is essential for the numerical integration of the geodesic equations in the next section.
	
	Once a numerical solution is obtained, we can extract the ADM mass and angular momentum from the asymptotic sub-leading behavior of the metric functions as $r\rightarrow\infty$
	\begin{equation}
		\begin{aligned}
			&g_{tt}=-1+\frac{2M}{r}+\mathcal{O}\left(\frac{1}{r^2}\right),\\
			&g_{t\varphi}=-\frac{2J}{r}\sin^2\theta+\mathcal{O}\left(\frac{1}{r^2}\right).
		\end{aligned}
	\end{equation}
	In practice, the ADM mass $M$ and angular momentum $J$ are obtained by
	\begin{equation}
		M=L\partial_x F_0\big|_{x=1},\;\;J=-\frac{L}{2}\partial_x \bar{W}\big|_{x=1},
	\end{equation}
	where $\bar{W}=W/r^2$. The parameter space of the parity-odd RBS, depicted in Figure 1 where a typical inspiralling pattern is shown over the frequency range $w_{\text{min}}<w<1$ with $w_{\text{min}}\approx0.6835$, has 13 marked points that serve as the typical cases to be discussed later in the paper, and the specific parameters of these points are presented in Table \ref{Table: Ref}.

	\begin{figure}
		\centering	
		\includegraphics[width=0.5\textwidth,height=0.425\textwidth]{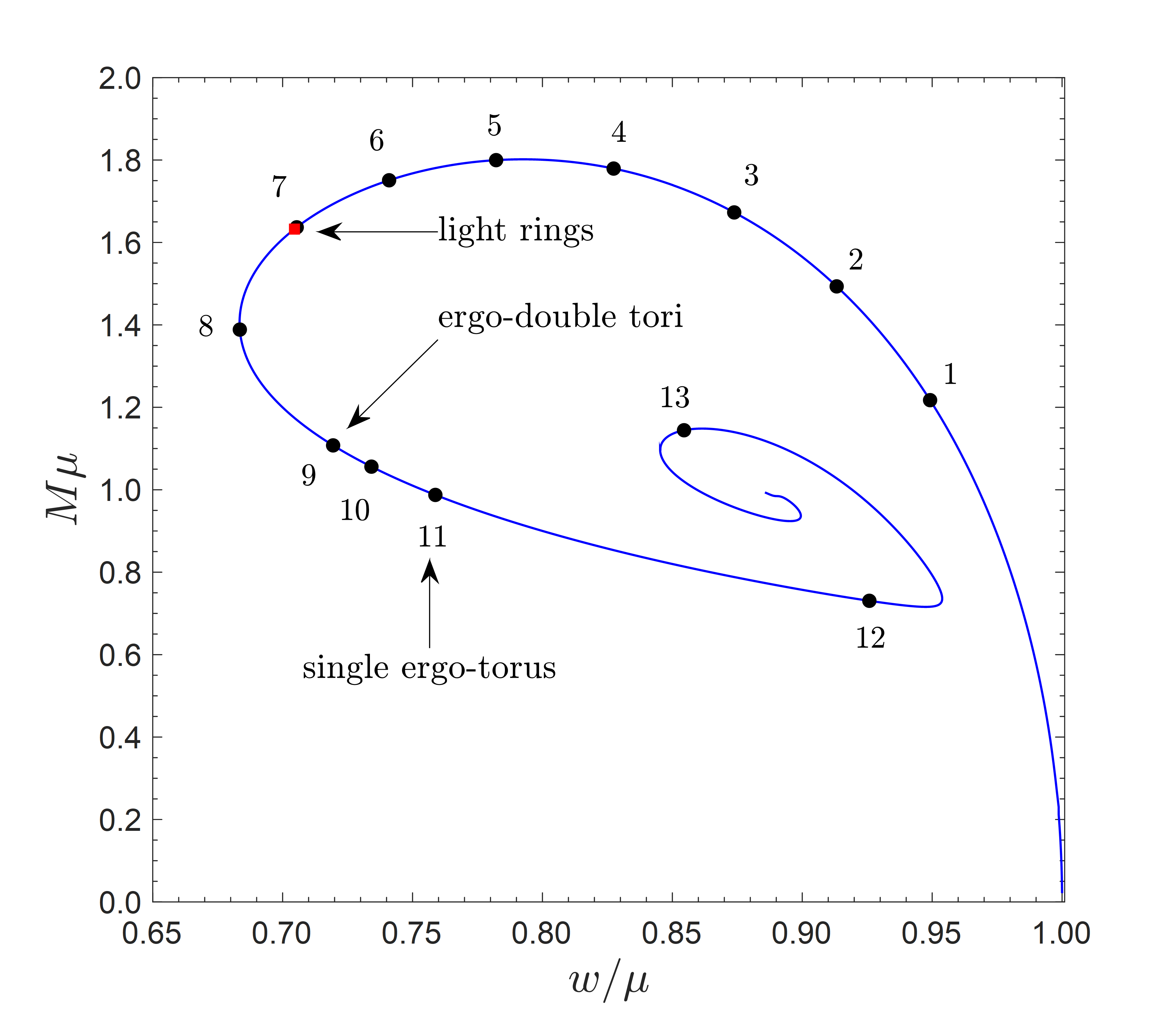}	
		\caption{Parity-odd RBS solutions in an Arnowitt-Deser-Misner (ADM) mass $M$ vs scalar field frequency $w$ diagram. Points $1$-$13$ corresponds to the BSs to be discussed.}
		\label{Fig: spiral}
	\end{figure}
	
	\begin{table}
		\begin{center}
			\caption{Reference Table of the BSs under discussion.}
			\label{Table: Ref}
			\begin{tabular}{l*{6}{c}r}
				\hline
				\hline
				& BS   &  $J$    & $M$     & $w$     & Light rings & Ergoregions     & \\
				\hline
				& $1$  & $1.237$ & $1.217$ & $0.949$ & No          & No              & \\
				\hline
				& $2$  & $1.533$ & $1.493$ & $0.913$ & No          & No          	 & \\
				\hline
				& $3$  & $1.734$ & $1.673$ & $0.874$ & No          & No          	 & \\
				\hline
				& $4$  & $1.858$ & $1.779$ & $0.827$ & No          & No          	 & \\
				\hline
				& $5$  & $1.883$ & $1.800$ & $0.782$ & No          & No          	 & \\
				\hline
				& $6$  & $1.819$ & $1.751$ & $0.741$ & No          & No          	 & \\
				\hline
				& $7$  & $1.660$ & $1.637$ & $0.705$ & No          & No          	 & \\
				\hline
				& $8$  & $1.301$ & $1.389$ & $0.684$ & $2$ stable $+$ & No           & \\
				&      &         &         &         & $2$ unstable&             	 & \\
				\hline
				& $9$  & $0.897$ & $1.108$ & $0.719$ & $2$ stable $+$ & Ergo-double-tori& \\
				&      &        &          &         & $2$ unstable&             	 & \\
				\hline
				& $10$ & $0.826$ & $1.056$ & $0.734$ & $2$ stable $+$ & Ergo-double-tori& \\
				&      &         &         &         & $2$ unstable&               & \\
				\hline
				& $11$ & $0.734$ & $0.988$ & $0.759$ & $2$ stable $+$ & Merger     & \\
				&      &         &         &         & $2$ unstable&               & \\
				\hline
				& $12$ & $0.423$ & $0.731$ & $0.926$ & $2$ stable $+$ & Single ergo-torus& \\
				&      &         &         &         & $2$ unstable&              	& \\
				\hline
				& $13$ & $0.870$ & $1.144$ & $0.854$ & $2$ stable $+$ & Single ergo-torus& \\
				&      &         &         &         & $2$ unstable&              	& \\
				\hline
				\hline
			\end{tabular}
		\end{center}
	\end{table}
	
	\section{Lensing and light rings}\label{Sec: Lensing}
	\subsection{Setup}
	In this section, we focus on the lensing effects of parity-odd RBSs. This reveals how the presence of BSs distorts the apparent sky. To investigate the lensing of parity-odd RBS, we employ a ray-tracing algorithm similar to \cite{Huang2018}. The geodesic equations are given by
	\begin{equation}
		\frac{dx^\alpha}{d\lambda}=p^\alpha,
	\end{equation}
	and
	\begin{equation}
		\frac{dp^\alpha}{d\lambda}=-\Gamma^\alpha_{\mu\nu}p^\mu p^\nu,
	\end{equation}
	The initial conditions of the light rays are specified by the observation angles seen by the observer. We follow a similar setup as described in \cite{Cunha:2015yba,Cunha:2016bjh,Cunha:2016bpi}. For instance, we assume the observer is located at $\tilde{r}=15M$, where the circumferential radius $\tilde{r}$ is given by \cite{Cunha:2016bpi}
	\begin{equation}
		\tilde{r}\equiv\frac{1}{2\pi}\int_{0}^{2\pi}\sqrt{g_{\varphi\varphi}}d\varphi.
	\end{equation}
	In the reference frame of zero angular momentum observers (ZAMO), the observation angles $(\alpha,\beta)$ can be related to the initial conditions of $p_\mu$ as follows:
	\begin{equation}
		\begin{aligned}
			&p_r      =|\vec{P}|\sqrt{g_{rr}}\cos\beta\cos\alpha,\\
			&p_\theta =|\vec{P}|\sqrt{g_{\theta\theta}}\sin\alpha,\\
			&p_\varphi=|\vec{P}|\sqrt{g_{\varphi\varphi}}\sin\beta\cos\alpha,\\
			&p_t      =|\vec{P}|\left(\frac{1+\gamma\sqrt{g_{\varphi\varphi}}\sin\beta\cos\alpha}{\zeta}\right),
		\end{aligned}
	\end{equation}
	where
	\begin{equation}
		\zeta=\sqrt{\frac{g_{\varphi\varphi}}{g^2_{t\varphi}-g_{tt}g_{\varphi\varphi}}},\;\;\gamma=-\frac{g_{t\varphi}}{g_{\varphi\varphi}}\zeta,
	\end{equation}
	and $|\vec{P}|$ only determines the photon's frequency and does not influence the trajectory. For simplicity, one can set it to unity. By definition, the initial conditions for $p^\alpha$ is given by $p^\alpha=g^{\alpha\beta}p_\beta$.
	
	In the integration of the geodesic equations, we employed two distinct methods: the 4th-order Runge-Kutta method (RK4) and the velocity Verlet algorithm \cite{Dolence:2009zz,Bronzwaer:2018lde}, both of which were implemented with adaptive step-size techniques. It turns out that the velocity Verlet algorithm is more robust for trajectories that pass through the poles at $\theta=0$ or $\pi$.
	
\begin{figure}
		\centering	
		\includegraphics[width=0.35\textwidth]{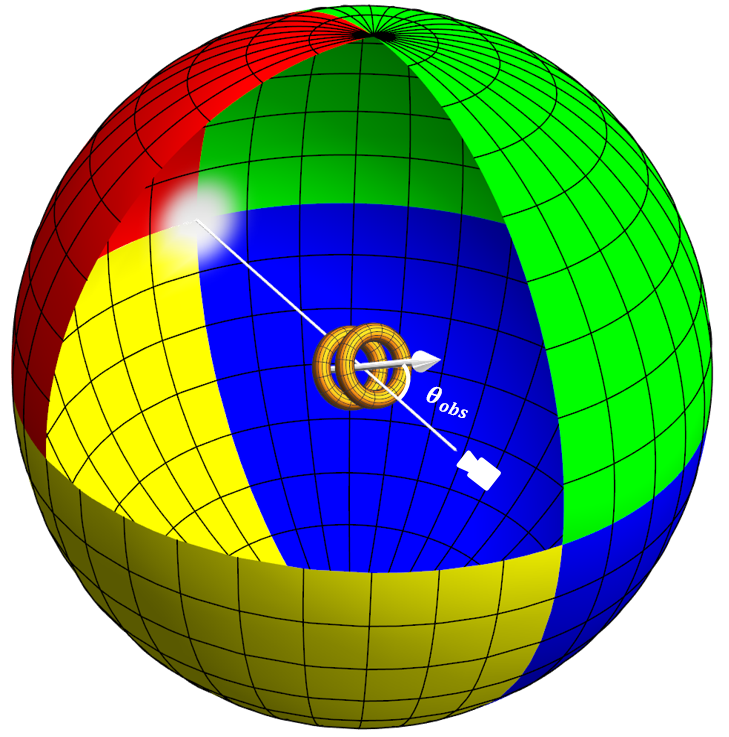}
		\caption{The full celestial sphere. The observer looks toward the center of the full celestial sphere, where the RBS is located. A white spot on the celestial sphere highlights the observer’s line of sight, while the white arrow indicates the axis of rotation of the RBS.}
		\label{Fig: celestial}
	\end{figure}

	To illustrate the distortion introduced by the RBSs, we divide the celestial sphere into four quadrants, each assigned a distinct color based on the angular coordinates on the sphere (see Fig. \ref{Fig: celestial}). Assuming the observer looks toward the center of the full celestial sphere, where the RBS is located, a white spot on the sphere highlights the observer’s line of sight, while a white arrow indicates the axis of rotation of the RBS, which lies in the plane of $0^\circ$ and $180^\circ$ of the celestial longitude coordinates. The angle between the white arrow and the observer’s line of sight is $\theta_{\text{obs}}$, which also represents the initial value of $\theta$ for the photons in the ray-tracing algorithm. Then, the lensing image is generated by connecting each pixel in the observer's screen and a point on the celestial sphere by the ray trajectory. This setup was introduced in \cite{Bohn2015} and has been widely used in \cite{Cunha:2016bjh,Cunha2017,Huang2018,Cunha:2015yba,Chen:2023wzv,Chen:2022scf,Sun:2023syd,Long:2020wqj,Wang:2018eui,Wang:2017qhh,Kuang:2024ugn,Hu:2020usx,Zhang:2022osx,Zhong:2021mty,Yang:2024ulu,Liu:2024lve,Liu:2024lbi,Liu:2024soc,Yuan:2024ltr,Zhang:2022osx,He:2022opa,Guo:2022muy}.
	
	\subsection{Lensing images}
\begin{figure*}
		\centering	
		\includegraphics[width=0.8\textwidth]{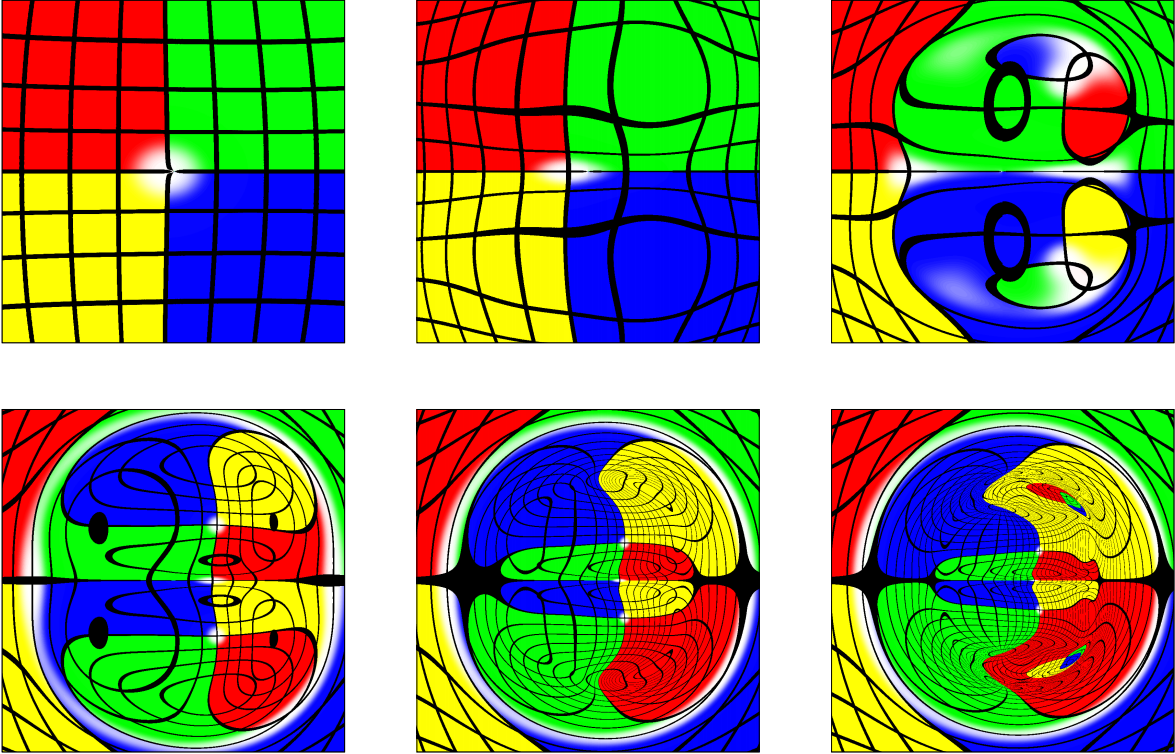}
		\caption{Lensing by parity-odd RBSs. From left to right: (top) $w_{1,2,3}=0.949;0.913;0.974$; (bottom) $w_{4,5,6}=0.827;0.782;0.741$.}
		\label{Fig: BSs low}
	\end{figure*}

\begin{figure*}
		\centering	
		\includegraphics[width=0.8\textwidth]{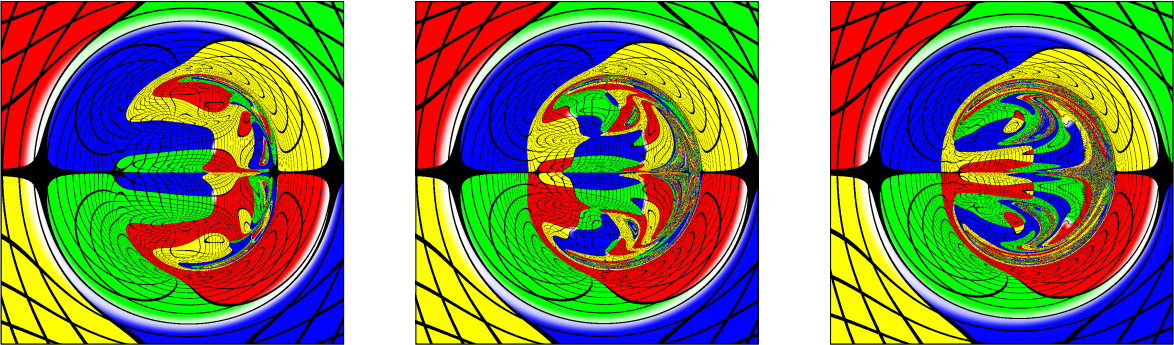}
		\caption{Lensing by parity-odd RBSs. From left to right: $w_{7,8,9}=0.705;0.684;0.719$.}
		\label{Fig: BSs mid}
	\end{figure*}

\begin{figure*}
		\centering	
		\includegraphics[width=0.8\textwidth]{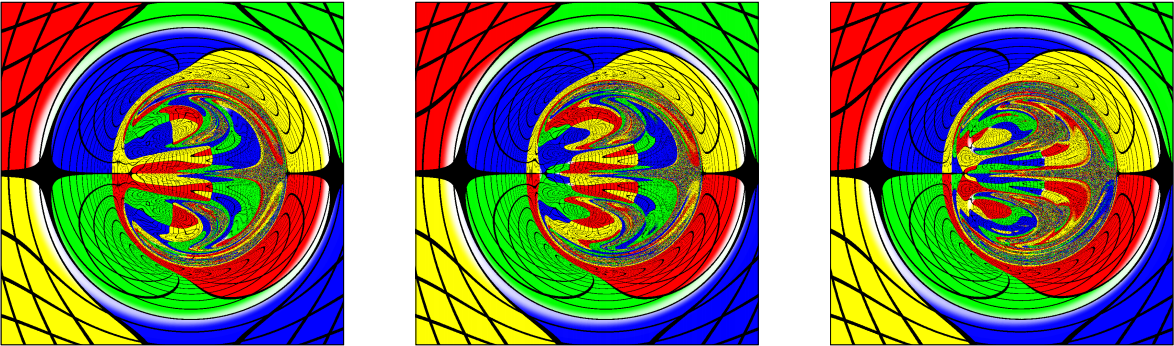}
		\caption{Lensing by ultra-compact parity-odd RBSs. From left to right: $w_{10,11,12}=0.734;0.759;0.926$.}
		\label{Fig: BSs ultra}
	\end{figure*}

\begin{figure*}
		\centering	
		\includegraphics[width=0.8\textwidth]{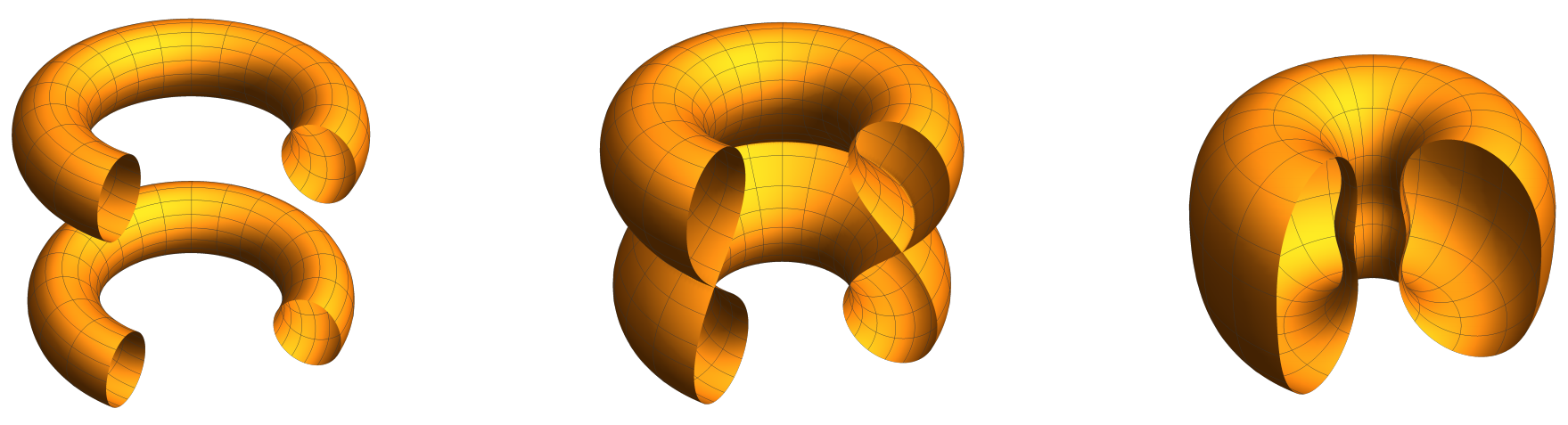}
		\caption{Ergoregions of RBS 10 (left), 11 (middle), 12 (right).}
		\label{Fig: ergoregions}
	\end{figure*}

\begin{figure}
		\centering	
		\includegraphics[width=0.325\textwidth]{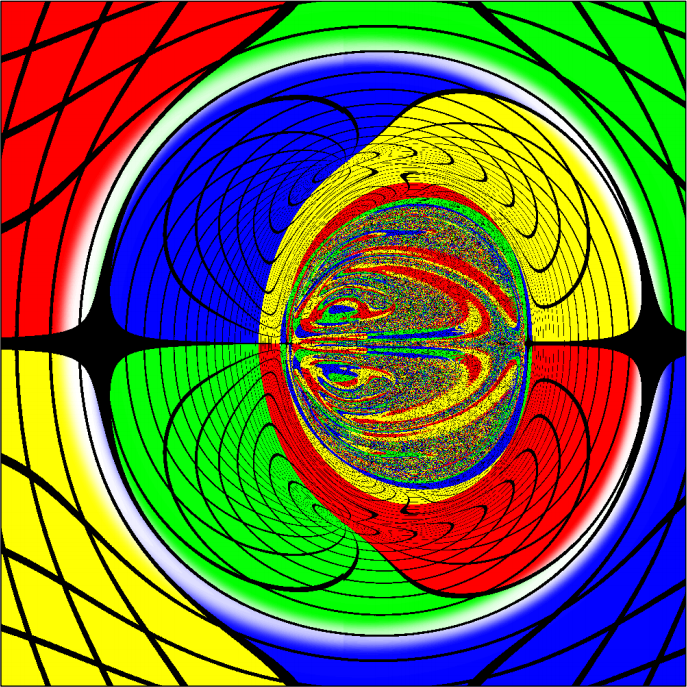}
		\caption{Lensing by ultra-compact RBS 13.}
		\label{Fig: BSs ultra2}
	\end{figure}

\begin{figure*}
		\centering	
		\includegraphics[width=0.9\textwidth]{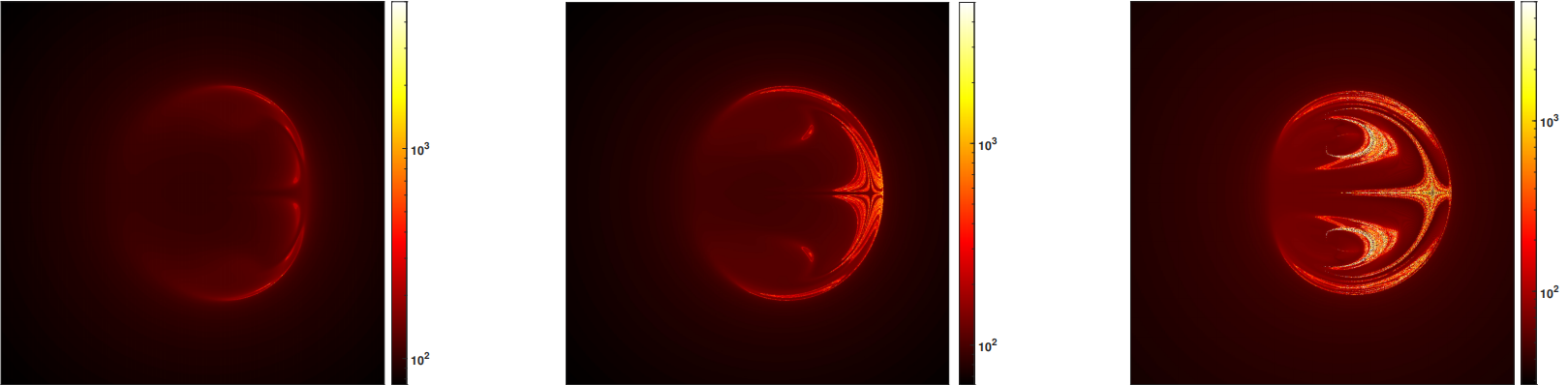}
		\caption{Time delay heat maps of RBSs 7 (left), 8 (middle), and 12 (right).}
		\label{Fig: BSs heat}
	\end{figure*}

\begin{figure*}
		\centering	
		\includegraphics[width=0.95\textwidth]{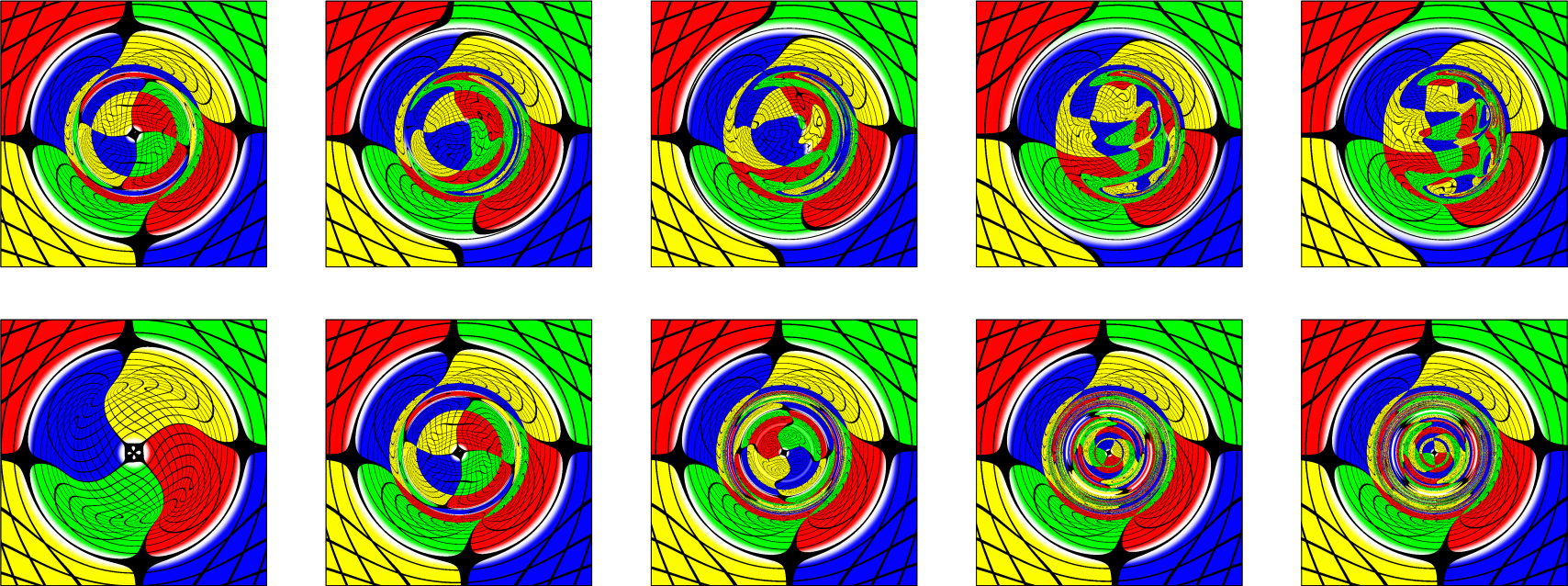}
		\caption{Lensing images for observer located away from the equatorial plane ($\theta_{\text{obs}}\neq 90^\circ$). Top panel, from left to right: images of RBS 7 for $\theta_{\text{obs}}=0^\circ,15^\circ,30^\circ,60^\circ $ and $75^\circ$; Bottom panel, from left to right: images at $\theta_{\text{obs}}=0^\circ$ for RBSs $5,7,8,9$ and $10$. }
		\label{Fig: BSs angle}
	\end{figure*}

	In this section, we present a collection of lensing images for parity-odd RBSs. Figures \ref{Fig: BSs low}-\ref{Fig: BSs heat} display the lensing images for observers located on the equatorial plane ($\theta_{\text{obs}}=\pi/2$). While Figure \ref{Fig: BSs angle} shows the lensing images for observer located away from the equatorial plane ($\theta_{\text{obs}}\neq\pi/2$).
	
	In Fig. \ref{Fig: BSs low}, we show the lensing images of RBSs 1-6. First, for RBS 1, there is no noticeable distortion in the image, as this solution is closest to the vacuum. As we proceed, the distortions caused by the RBSs are more apparent and clearer. For RBS 2, the white spot has an elliptical shape, while for RBS 3, the white spot is split into several disconnected parts, which are multiple images of a point source due to gravitational lensing. Next, for RBS 4 and 5, the Einstein ring appears, and we can see three copies of the white spot inside the ring, all of which are elliptically shaped. We also observe complete copies of the four quadrants of the celestial sphere. Moving on to RBS 6, the distortions become even more pronounced, showing multi-copies for the first time in the results.
	
	From Fig. \ref{Fig: BSs low}, it is observed that there is a general characteristic for parity-odd RBSs: when Einstein rings are present, the copies of the four quadrants of the celestial sphere consistently exhibit twice as much as those for parity-even RBSs described in \cite{Cunha:2015yba}. This behavior is evident only in plots with Einstein rings, and supports the argument that parity-odd RBSs can be interpreted as two RBSs in equilibrium \cite{Herdeiro:2023roz}.
	
	Figure \ref{Fig: BSs mid} shows the lensing images of RBSs 7-9. RBSs 7 and 9 represent two turning points in the parameter space: RBS 7 is very close to the solution where light rings first appear, while RBS 9 marks the first appearance of an ergoregion. Therefore, all solutions between RBSs 7 and 9 possess light rings but no ergoregions. Thus, RBS with an ergoregion must have light rings outside \cite{Ghosh:2021txu}. Focusing on the lensing images in this figure, we observe an infinite number of copies of the four quadrants of the celestial sphere, resulting in chaotic patterns within the Einstein ring. The existence of light rings may make it difficult for photons to escape, creating pockets in the effective potential that can trap photons for longer periods \cite{Cunha:2016bjh}. In the next section, we will analyze the effective potential of some solutions and show that this statement is also true for odd-parity BSs.
	
	Moving on to Fig. \ref{Fig: BSs ultra}, we turn our attention to the lensing effects of RBSs 10-12, all of which possess ergoregions and four light rings. The detailed structure of these ergoregions is presented in Fig. \ref{Fig: ergoregions}. Similar to Fig. \ref{Fig: BSs mid}, the lensing images of RBSs 10-12 also exhibit chaotic patterns within the Einstein ring, with the chaotic region becoming more pronounced as we move from RBSs 10 to 12. According to \cite{Cunha:2016bjh}, the presence of stable light rings might contribute to these chaotic patterns. All RBSs in this figure have ergoregions and four light rings (two stable and two unstable).
	
	For BS solutions that move more inward along the spiral curve in Fig. \ref{Fig: spiral}, numerical calculations become increasingly challenging. To maintain numerical accuracy, we must decrease the value of $L$ in Eq. (\ref{Eq: copact coord}). Typically, we set $L=0.25$ for numerical computations in these more inward regions. Despite the increased challenges, no unusual effects were observed in the lensing images for RBSs more inward than RBS 12 along the spiral curve in Fig. \ref{Fig: spiral}. A general observation is that the chaotic regions in the lensing images became considerably larger, as shown in Fig. \ref{Fig: BSs ultra2}, which displays lensing image of RBS 13.
	
	The time delay function, defined as the variation of the coordinate time required for a photon emitted from a particular pixel to reach the celestial sphere, can be used to understand the chaotic lensing image of the RBSs \cite{Cunha:2016bjh}. In Fig. \ref{Fig: BSs heat}, we plot the time delay function of RBSs 7, 8, and 12. We can see clearly that the brighter regions on the time delay image correspond closely with the chaotic regions in the lensing image.
	
	Now, let us consider the lensing images for observers located away from the equatorial plane, as illustrated in Fig. \ref{Fig: BSs angle}. There are many interesting observations in this figure. First, we can clearly observe the frame dragging caused by the spin of the BSs: the space is dragged in the direction of the rotation, see especially the bottom panel. Second, we can see Einstein rings originated from the white reference spot, as well as the inverted image inside the ring. In the leftmost image of the bottom panel in Fig. \ref{Fig: BSs angle}, one can observe only one Einstein ring and the corresponding inverted image. Moving to the right, as shown in the subsequent images, the RBS becomes increasingly compact. Consequently, multiple Einstein rings and inverted images begin to appear. Furthermore, when the BS becomes increasingly compact, chaotic lensing phenomena also become apparent, as seen in the middle image of the bottom panel.
	
\begin{figure*}
		\centering	
		\includegraphics[width=0.8\textwidth]{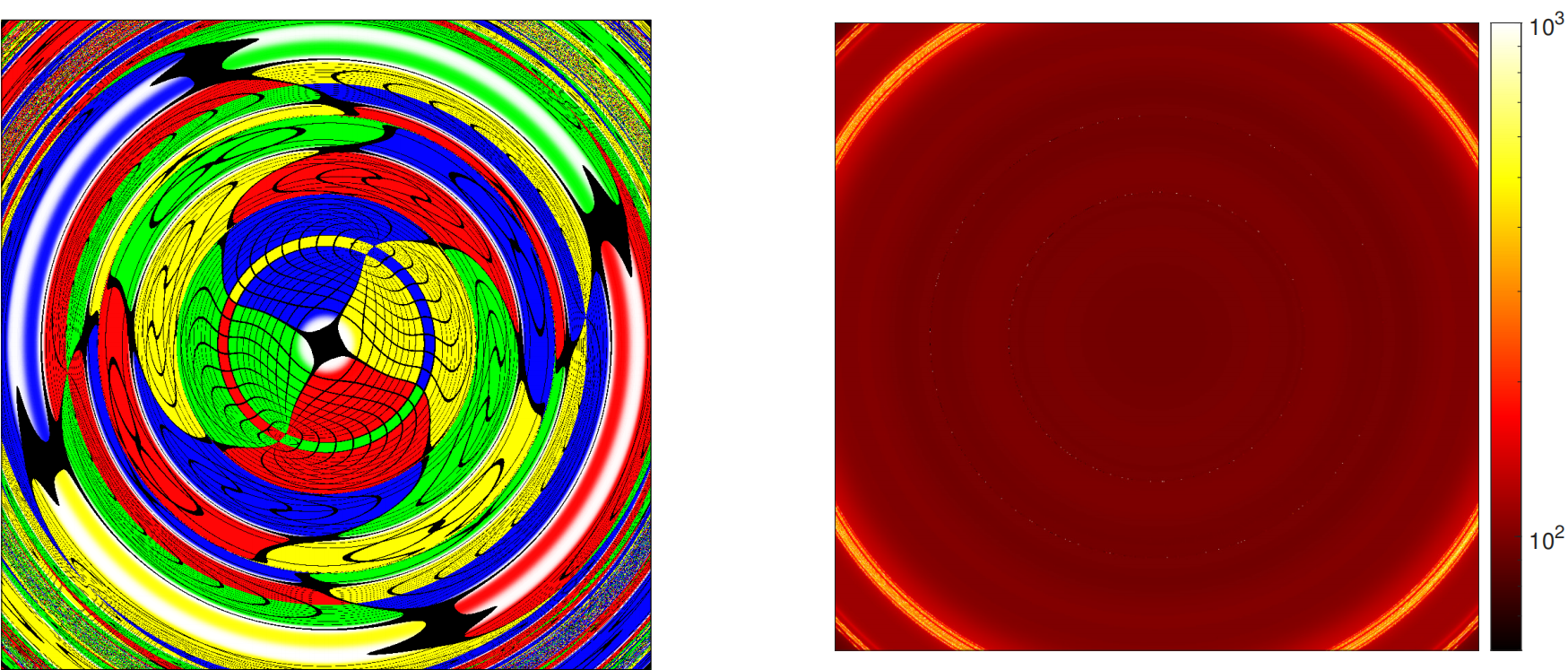}
		\caption{A zoom of the lensing image of RBS 9 (left panel), and the corresponding time delay function (right panel). }
		\label{Fig: BSs angle in}
	\end{figure*}

	In Fig. \ref{Fig: BSs angle in}, we present a zoom-in view of the lensing image of RBS 9 at $\theta_{\text{obs}}=0^\circ$ along with the corresponding time delay function. The left panel shows a multi-Einstein ring structure within the strong field region, along with chaotic patterns that exhibit a similar multi-ring structure. In the right panel, we observe that the brighter regions in the time delay function also form a multi-ring structure, which appears to coincide with the chaotic regions observed in the left panel. This can be interpreted as the multi-image of the light rings of the parity-odd RBS.
	
	\subsection{Light rings}
	In this section, we will consider the properties of the light rings for the parity-odd RBSs. Following \cite{Cunha:2016bjh}, we write the Hamiltonian in the form
	\begin{equation}
		2\mathcal{H}=T+V=0,
	\end{equation}
	where
	\begin{equation}
		T=p^2_{r}g^{rr}+p^2_{\theta}g^{\theta\theta}\geq0,
	\end{equation}
	is the kinetic energy and the potential is given by
	\begin{equation}
		V=p^2_tg^{tt}+p^2_{\varphi} g^{\varphi\varphi}+2p_t p_{\varphi} g^{t\varphi}\leq0.
	\end{equation}
	This inequality defines the allowed region in the $(r,\theta)$-space. For the rotating BSs, all the metric components depend only on $(r,\theta)$. Consequently, there are two Killing vectors, $\partial_t$ and $\partial_\varphi$, which lead to two conserved quantities
	\begin{equation}
		E=-p_t,\;\;\;\Phi=p_{\varphi}.
	\end{equation}
	Defining the impact factor $\eta\equiv \Phi/E$, one can rewrite this inequality as (with $g_{tt}\neq0$)
	\begin{equation}\label{Eq: potential}
		V=-\frac{E^2}{D}g_{tt}\left(\eta-h_+\right)\left(\eta-h_-\right)\leq0,
	\end{equation}
	where
	\begin{equation}
		h_{\pm}=\frac{-g_{t\varphi}\pm\sqrt{D}}{g_{tt}},
	\end{equation}
	and
	\begin{equation}
		D=g^2_{t\varphi}-g_{tt}g_{\varphi\varphi}=r^2\sin^2\theta e^{2(F_0+F_2)}\geq0.
	\end{equation}
	
	The existence of light rings is governed by the potential function $V(r,\theta)$ (or $h_{\pm}(r,\theta)$). By definition, light rings refer to geodesic with $p_r=p_{\theta}=0$ and $\dot{p}_r=\dot{p}_\theta=0$, which lead to
	\begin{equation}
		\eta-h_{\pm}=\partial_r h_{\pm}=\partial_\theta h_{\pm}=0.
	\end{equation}
	We find that for parity-odd RBSs, stable light rings are always associated with the local minima of the effective potential $h_+$, while unstable ones correspond to saddle points in the potential.
	
\begin{figure*}
		\centering	
		\includegraphics[width=0.8\textwidth]{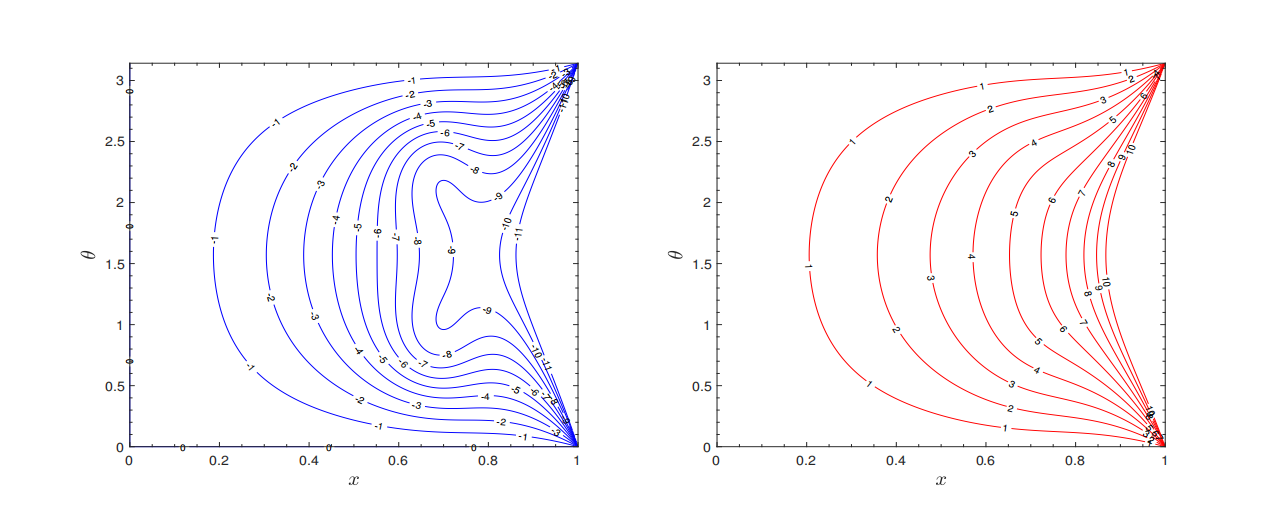}
		\caption{Contour plots of $h_+$ (left) and $h_-$ (right) for RBS 7. It has no ergoregion or unstable light rings, but it has two stable light rings symmetrically distributed about the equatorial plane.}
		\label{Fig: hp sol7}
	\end{figure*}

	In Fig. \ref{Fig: hp sol7}, we present the contour plots of $h_{\pm}$ for RBS 6. Since $h_-$ exhibits a similar structure across all cases, subsequent figures will focus only on $h_+$. From the left panel of Fig. \ref{Fig: hp sol7}, we can clearly observe that two pockets are about to form in $h_+$. These pockets, which are expected to appear symmetrically on either side of the equatorial plane, are set to close below a certain value of $\eta$ and create an allowed region that will be inaccessible from spatial infinity. 	
	
\begin{figure}
		\centering	
		\includegraphics[width=0.45\textwidth]{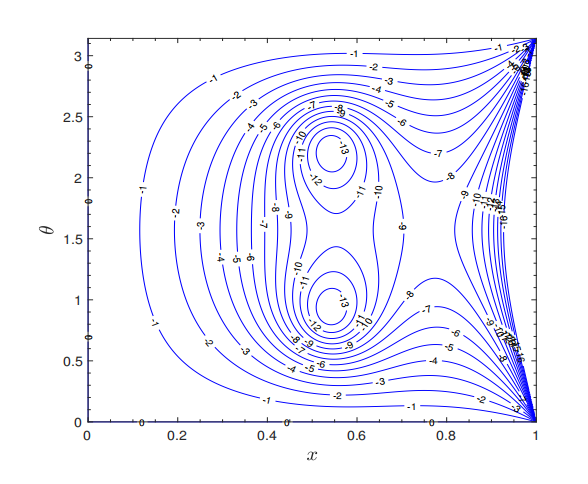}	
		\caption{Contour plots of $h_+$ for RBS 8. It has no ergoregion but possesses two stable light rings symmetrically distributed about the equatorial plane, along with two unstable light rings located precisely on the equatorial plane.}
		\label{Fig: hp sol8}
	\end{figure}

	Figure \ref{Fig: hp sol8} displays the contour plot of $h_+$ for RBS 8. It is observed that the pockets can be closed for $\eta<-9$, forming an allowed region that is disconnected from spatial infinity. Furthermore, when $\eta<-11$, additional closed contours emerge in pairs, symmetrically distributed around the equatorial plane, which correspond to two disjoint allowed regions for particle motion, each localized on either side of the equatorial plane. Finally, as the impact parameter is further decreased to $\eta\approx-13.78$, the two disjoint allowed regions become arbitrarily small, ultimately shrinking to two points symmetric about the equatorial plane, which represent the minima of $h_+$ and correspond to two stable photon rings that are symmetrically positioned relative to the equatorial plane.
	
\begin{figure*}
		\centering	
		\includegraphics[width=0.8\textwidth]{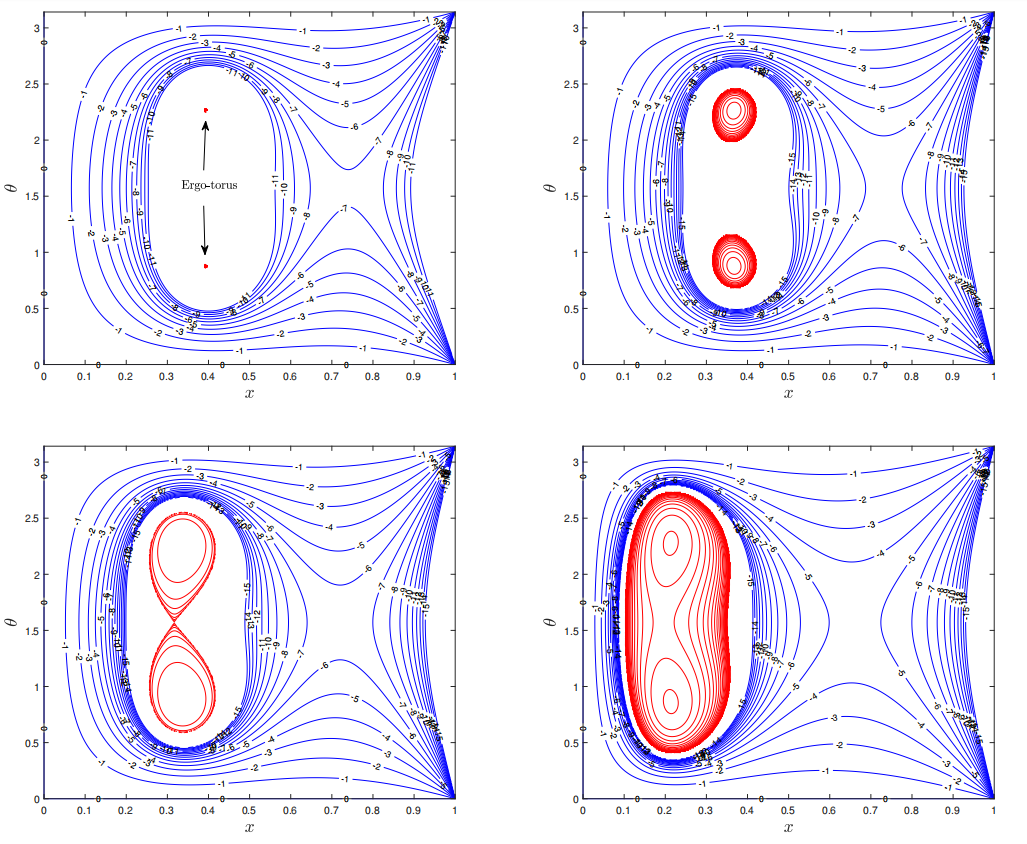}			
		\caption{Contour plots of $h_+$ for RBSs 9-12. From left to right (top): RBSs 9 and 10; (bottom): RBSs 11 and 12. The red lines represent the positive values of $h_+$, while the blue lines represent negative values of $h_+$.}
		\label{Fig: hp sol9}
	\end{figure*}

	In Fig. \ref{Fig: hp sol9}, we show the contour plot of $h_+$ for RBSs with ergoregions. In these cases, the potential $h_+$ exhibits divergence as the ergosurface is approached, whether from within or beyond the ergoregions. Specifically, when approached from the interior, $h_+$ tends to $+\infty$ from a positive minimum, which corresponds to stable light rings located inside the ergoregion. Conversely, when approached from the exterior, $h_+$ remains negative and diverges to $-\infty$ at the ergosurface. Both RBSs 9 and 10 have two saddle points on the equatorial plane; however, one of them is not shown in the plot, as $h_+$ decreases sharply near the ergoregions. RBS 11 marks a critical point in the parameter space, where the double ergo-tori are about to merge into a single ergo-torus, with one saddle point of $h_+$ poised to enter the ergoregion. For RBSs more compact than RBS 11, the double ergo-tori have fully merged into a single ergo-torus, and one of the saddle points of $h_+$ has indeed moved into the ergoregion, as clearly depicted in the bottom right panel of Fig. \ref{Fig: hp sol9}, where we see clearly that for RBS 12, the $h_+$ function within the ergoregion has two local minima and one saddle point, corresponding to two stable light rings and one unstable light ring inside the ergoregion, respectively. Outside the ergoregion, there is a single saddle point, corresponding to an unstable light ring.
	
	\section{Conclusion}\label{Sec: conclude}
	In this paper, we have carried out a comprehensive analysis of the gravitational lensing and light rings associated with parity-odd RBS solutions. Our findings reveal significant characteristics and behaviors that distinguish parity-odd RBSs from their parity-even counterparts. The main results are summarized as follows.
	
	First, we analyzed the lensing images of parity-odd RBSs for an observer located at the equatorial plane. We showed that an Einstein ring along with an inverted image will appear when the BS becomes very compact. Our images revealed that the number of replicated inverted images due to the lensing effect is always twice the number of images for parity-even counterparts, see Figs. \ref{Fig: BSs low}-\ref{Fig: BSs ultra}.
	
	Next, we considered the lensing images for observers located away from the equatorial plane. Our results showed the frame-dragging effect caused by the spin of the BSs. We also demonstrated that both the Einstein rings and the brighter regions in the time delay function, which are related to the stable light rings, exhibit a multi-ring structure.
	
	Finally, we analyzed the light rings of parity-odd RBSs. We showed that the stable light rings, if they exist, are always symmetrically distributed about the equatorial plane, while the unstable light rings are located exactly on the equatorial plane. Moreover, the number of light rings for parity-odd BSs is twice that of the parity-even ones. Our results also showed that for BSs possessing light rings, there are always chaotic patterns in the lensing image.
	
	Although our study has provided valuable insights into the lensing and light rings of parity-odd RBSs, future work could extend this analysis to more realistic scenarios. Specifically, considering the imaging of bosonic stars with more realistic light sources, such as accretion disks \cite{Meng:2023uws,Wang:2023fge,Qu:2023hsy}, would be an intriguing avenue for further research. Such investigations could offer deeper understanding of the complexities and unique features of bosonic stars in astrophysical contexts.
	
	\begin{acknowledgments}
		This work was supported by the National Natural Science Foundation of China (Grants Nos. 12275106 and 12235019), the Shandong Provincial Natural Science Foundation (Grant No. ZR2024QA032), and the Innovation Grogram of Shanghai Normal University (Grant No. KF202472). Some of the calculations were performed in High-performance Computing Platform at University of Jinan. YH would like to thank Professor Carlos Herdeiro for his prompt and helpful correspondence.
	\end{acknowledgments}
	
	\bibliography{boson.bib}

\begin{thebibliography}{76}%
\makeatletter
\providecommand \@ifxundefined [1]{%
 \@ifx{#1\undefined}
}%
\providecommand \@ifnum [1]{%
 \ifnum #1\expandafter \@firstoftwo
 \else \expandafter \@secondoftwo
 \fi
}%
\providecommand \@ifx [1]{%
 \ifx #1\expandafter \@firstoftwo
 \else \expandafter \@secondoftwo
 \fi
}%
\providecommand \natexlab [1]{#1}%
\providecommand \enquote  [1]{``#1''}%
\providecommand \bibnamefont  [1]{#1}%
\providecommand \bibfnamefont [1]{#1}%
\providecommand \citenamefont [1]{#1}%
\providecommand \href@noop [0]{\@secondoftwo}%
\providecommand \href [0]{\begingroup \@sanitize@url \@href}%
\providecommand \@href[1]{\@@startlink{#1}\@@href}%
\providecommand \@@href[1]{\endgroup#1\@@endlink}%
\providecommand \@sanitize@url [0]{\catcode `\\12\catcode `\$12\catcode
  `\&12\catcode `\#12\catcode `\^12\catcode `\_12\catcode `\%12\relax}%
\providecommand \@@startlink[1]{}%
\providecommand \@@endlink[0]{}%
\providecommand \url  [0]{\begingroup\@sanitize@url \@url }%
\providecommand \@url [1]{\endgroup\@href {#1}{\urlprefix }}%
\providecommand \urlprefix  [0]{URL }%
\providecommand \Eprint [0]{\href }%
\providecommand \doibase [0]{http://dx.doi.org/}%
\providecommand \selectlanguage [0]{\@gobble}%
\providecommand \bibinfo  [0]{\@secondoftwo}%
\providecommand \bibfield  [0]{\@secondoftwo}%
\providecommand \translation [1]{[#1]}%
\providecommand \BibitemOpen [0]{}%
\providecommand \bibitemStop [0]{}%
\providecommand \bibitemNoStop [0]{.\EOS\space}%
\providecommand \EOS [0]{\spacefactor3000\relax}%
\providecommand \BibitemShut  [1]{\csname bibitem#1\endcsname}%
\let\auto@bib@innerbib\@empty
\bibitem [{\citenamefont {Wheeler}(1955)}]{Wheeler:1955zz}%
  \BibitemOpen
  \bibfield  {author} {\bibinfo {author} {\bibfnamefont {J.~A.}\ \bibnamefont
  {Wheeler}},\ }\href {\doibase 10.1103/PhysRev.97.511} {\bibfield  {journal}
  {\bibinfo  {journal} {Phys. Rev.}\ }\textbf {\bibinfo {volume} {97}},\
  \bibinfo {pages} {511} (\bibinfo {year} {1955})}\BibitemShut {NoStop}%
\bibitem [{\citenamefont {Power}\ and\ \citenamefont
  {Wheeler}(1957)}]{Wheeler:1957}%
  \BibitemOpen
  \bibfield  {author} {\bibinfo {author} {\bibfnamefont {E.~A.}\ \bibnamefont
  {Power}}\ and\ \bibinfo {author} {\bibfnamefont {J.~A.}\ \bibnamefont
  {Wheeler}},\ }\href {\doibase 10.1103/RevModPhys.29.480} {\bibfield
  {journal} {\bibinfo  {journal} {Rev. Mod. Phys.}\ }\textbf {\bibinfo {volume}
  {29}},\ \bibinfo {pages} {480} (\bibinfo {year} {1957})}\BibitemShut
  {NoStop}%
\bibitem [{\citenamefont {Kaup}(1968)}]{Kaup1968}%
  \BibitemOpen
  \bibfield  {author} {\bibinfo {author} {\bibfnamefont {D.~J.}\ \bibnamefont
  {Kaup}},\ }\href {\doibase 10.1103/PhysRev.172.1331} {\bibfield  {journal}
  {\bibinfo  {journal} {Phys. Rev.}\ }\textbf {\bibinfo {volume} {172}},\
  \bibinfo {pages} {1331} (\bibinfo {year} {1968})}\BibitemShut {NoStop}%
\bibitem [{\citenamefont {Ruffini}\ and\ \citenamefont
  {Bonazzola}(1969)}]{Ruffini:1969qy}%
  \BibitemOpen
  \bibfield  {author} {\bibinfo {author} {\bibfnamefont {R.}~\bibnamefont
  {Ruffini}}\ and\ \bibinfo {author} {\bibfnamefont {S.}~\bibnamefont
  {Bonazzola}},\ }\href {\doibase 10.1103/PhysRev.187.1767} {\bibfield
  {journal} {\bibinfo  {journal} {Phys. Rev.}\ }\textbf {\bibinfo {volume}
  {187}},\ \bibinfo {pages} {1767} (\bibinfo {year} {1969})}\BibitemShut
  {NoStop}%
\bibitem [{\citenamefont {Jetzer}(1992)}]{Jetzer1992}%
  \BibitemOpen
  \bibfield  {author} {\bibinfo {author} {\bibfnamefont {P.}~\bibnamefont
  {Jetzer}},\ }\href {\doibase 10.1016/0370-1573(92)90123-h} {\bibfield
  {journal} {\bibinfo  {journal} {Physics Reports}\ }\textbf {\bibinfo {volume}
  {220}},\ \bibinfo {pages} {163} (\bibinfo {year} {1992})}\BibitemShut
  {NoStop}%
\bibitem [{\citenamefont {Schunck}\ and\ \citenamefont
  {Mielke}(2003)}]{Schunck2003}%
  \BibitemOpen
  \bibfield  {author} {\bibinfo {author} {\bibfnamefont {F.~E.}\ \bibnamefont
  {Schunck}}\ and\ \bibinfo {author} {\bibfnamefont {E.~W.}\ \bibnamefont
  {Mielke}},\ }\href {\doibase 10.1088/0264-9381/20/20/201} {\bibfield
  {journal} {\bibinfo  {journal} {Classical and Quantum Gravity}\ }\textbf
  {\bibinfo {volume} {20}},\ \bibinfo {pages} {R301} (\bibinfo {year}
  {2003})}\BibitemShut {NoStop}%
\bibitem [{\citenamefont {Visinelli}(2021)}]{Visinelli2021}%
  \BibitemOpen
  \bibfield  {author} {\bibinfo {author} {\bibfnamefont {L.}~\bibnamefont
  {Visinelli}},\ }\href {\doibase 10.1142/s0218271821300068} {\bibfield
  {journal} {\bibinfo  {journal} {Int. J. Mod. Phys. D Vol. 30, Issue 15, No.
  2130006 (2021)}\ }\textbf {\bibinfo {volume} {30}} (\bibinfo {year} {2021}),\
  10.1142/s0218271821300068},\ \Eprint {http://arxiv.org/abs/2109.05481}
  {arXiv:2109.05481 [gr-qc]} \BibitemShut {NoStop}%
\bibitem [{\citenamefont {Liebling}\ and\ \citenamefont
  {Palenzuela}(2023)}]{Liebling2023}%
  \BibitemOpen
  \bibfield  {author} {\bibinfo {author} {\bibfnamefont {S.~L.}\ \bibnamefont
  {Liebling}}\ and\ \bibinfo {author} {\bibfnamefont {C.}~\bibnamefont
  {Palenzuela}},\ }\href {\doibase 10.1007/s41114-023-00043-4} {\bibfield
  {journal} {\bibinfo  {journal} {Living Reviews in Relativity}\ }\textbf
  {\bibinfo {volume} {26}} (\bibinfo {year} {2023}),\
  10.1007/s41114-023-00043-4}\BibitemShut {NoStop}%
\bibitem [{\citenamefont {Khlopov}\ \emph {et~al.}(1985)\citenamefont
  {Khlopov}, \citenamefont {Malomed}, \citenamefont {Zeldovich},\ and\
  \citenamefont {Zeldovich}}]{Khlopov:1985fch}%
  \BibitemOpen
  \bibfield  {author} {\bibinfo {author} {\bibfnamefont {M.~Y.}\ \bibnamefont
  {Khlopov}}, \bibinfo {author} {\bibfnamefont {B.~A.}\ \bibnamefont
  {Malomed}}, \bibinfo {author} {\bibfnamefont {I.~B.}\ \bibnamefont
  {Zeldovich}}, \ and\ \bibinfo {author} {\bibfnamefont {Y.~B.}\ \bibnamefont
  {Zeldovich}},\ }\href {\doibase 10.1093/mnras/215.4.575} {\bibfield
  {journal} {\bibinfo  {journal} {Mon. Not. Roy. Astron. Soc.}\ }\textbf
  {\bibinfo {volume} {215}},\ \bibinfo {pages} {575} (\bibinfo {year}
  {1985})}\BibitemShut {NoStop}%
\bibitem [{\citenamefont {Guo}\ and\ \citenamefont
  {Zhang}(2019)}]{Guo:2018yyt}%
  \BibitemOpen
  \bibfield  {author} {\bibinfo {author} {\bibfnamefont {J.-Q.}\ \bibnamefont
  {Guo}}\ and\ \bibinfo {author} {\bibfnamefont {H.}~\bibnamefont {Zhang}},\
  }\href {\doibase 10.1140/epjc/s10052-019-7144-2} {\bibfield  {journal}
  {\bibinfo  {journal} {Eur. Phys. J. C}\ }\textbf {\bibinfo {volume} {79}},\
  \bibinfo {pages} {625} (\bibinfo {year} {2019})},\ \Eprint
  {http://arxiv.org/abs/1808.09826} {arXiv:1808.09826 [gr-qc]} \BibitemShut
  {NoStop}%
\bibitem [{\citenamefont {Guo}\ \emph {et~al.}(2020)\citenamefont {Guo},
  \citenamefont {Zhang}, \citenamefont {Chen}, \citenamefont {Joshi},\ and\
  \citenamefont {Zhang}}]{Guo:2020ked}%
  \BibitemOpen
  \bibfield  {author} {\bibinfo {author} {\bibfnamefont {J.-Q.}\ \bibnamefont
  {Guo}}, \bibinfo {author} {\bibfnamefont {L.}~\bibnamefont {Zhang}}, \bibinfo
  {author} {\bibfnamefont {Y.}~\bibnamefont {Chen}}, \bibinfo {author}
  {\bibfnamefont {P.~S.}\ \bibnamefont {Joshi}}, \ and\ \bibinfo {author}
  {\bibfnamefont {H.}~\bibnamefont {Zhang}},\ }\href {\doibase
  10.1140/epjc/s10052-020-08486-7} {\bibfield  {journal} {\bibinfo  {journal}
  {Eur. Phys. J. C}\ }\textbf {\bibinfo {volume} {80}},\ \bibinfo {pages} {924}
  (\bibinfo {year} {2020})},\ \Eprint {http://arxiv.org/abs/2011.06792}
  {arXiv:2011.06792 [gr-qc]} \BibitemShut {NoStop}%
\bibitem [{\citenamefont {Hu}\ \emph {et~al.}(2023)\citenamefont {Hu},
  \citenamefont {Guo}, \citenamefont {Li}, \citenamefont {Shao},\ and\
  \citenamefont {Zhang}}]{Hu:2023qcq}%
  \BibitemOpen
  \bibfield  {author} {\bibinfo {author} {\bibfnamefont {Y.}~\bibnamefont
  {Hu}}, \bibinfo {author} {\bibfnamefont {J.-Q.}\ \bibnamefont {Guo}},
  \bibinfo {author} {\bibfnamefont {J.}~\bibnamefont {Li}}, \bibinfo {author}
  {\bibfnamefont {C.-G.}\ \bibnamefont {Shao}}, \ and\ \bibinfo {author}
  {\bibfnamefont {H.}~\bibnamefont {Zhang}},\ }\href {\doibase
  10.1088/1572-9494/ad0029} {\bibfield  {journal} {\bibinfo  {journal} {Commun.
  Theor. Phys.}\ }\textbf {\bibinfo {volume} {75}},\ \bibinfo {pages} {125402}
  (\bibinfo {year} {2023})},\ \Eprint {http://arxiv.org/abs/2405.06351}
  {arXiv:2405.06351 [gr-qc]} \BibitemShut {NoStop}%
\bibitem [{\citenamefont {Mielke}\ and\ \citenamefont
  {Scherzer}(1981)}]{Mielke1981}%
  \BibitemOpen
  \bibfield  {author} {\bibinfo {author} {\bibfnamefont {E.~W.}\ \bibnamefont
  {Mielke}}\ and\ \bibinfo {author} {\bibfnamefont {R.}~\bibnamefont
  {Scherzer}},\ }\href {\doibase 10.1103/physrevd.24.2111} {\bibfield
  {journal} {\bibinfo  {journal} {Physical Review D}\ }\textbf {\bibinfo
  {volume} {24}},\ \bibinfo {pages} {2111} (\bibinfo {year}
  {1981})}\BibitemShut {NoStop}%
\bibitem [{\citenamefont {Colpi}\ \emph {et~al.}(1986)\citenamefont {Colpi},
  \citenamefont {Shapiro},\ and\ \citenamefont {Wasserman}}]{Colpi:1986ye}%
  \BibitemOpen
  \bibfield  {author} {\bibinfo {author} {\bibfnamefont {M.}~\bibnamefont
  {Colpi}}, \bibinfo {author} {\bibfnamefont {S.~L.}\ \bibnamefont {Shapiro}},
  \ and\ \bibinfo {author} {\bibfnamefont {I.}~\bibnamefont {Wasserman}},\
  }\href {\doibase 10.1103/PhysRevLett.57.2485} {\bibfield  {journal} {\bibinfo
   {journal} {Phys. Rev. Lett.}\ }\textbf {\bibinfo {volume} {57}},\ \bibinfo
  {pages} {2485} (\bibinfo {year} {1986})}\BibitemShut {NoStop}%
\bibitem [{\citenamefont {Lee}(1987)}]{Lee1987}%
  \BibitemOpen
  \bibfield  {author} {\bibinfo {author} {\bibfnamefont {T.~D.}\ \bibnamefont
  {Lee}},\ }\href {\doibase 10.1103/physrevd.35.3637} {\bibfield  {journal}
  {\bibinfo  {journal} {Physical Review D}\ }\textbf {\bibinfo {volume} {35}},\
  \bibinfo {pages} {3637} (\bibinfo {year} {1987})}\BibitemShut {NoStop}%
\bibitem [{\citenamefont {Li}\ \emph {et~al.}(2001)\citenamefont {Li},
  \citenamefont {Hao}, \citenamefont {Liu},\ and\ \citenamefont
  {Chen}}]{Li:2001he}%
  \BibitemOpen
  \bibfield  {author} {\bibinfo {author} {\bibfnamefont {X.-Z.}\ \bibnamefont
  {Li}}, \bibinfo {author} {\bibfnamefont {J.-G.}\ \bibnamefont {Hao}},
  \bibinfo {author} {\bibfnamefont {D.-J.}\ \bibnamefont {Liu}}, \ and\
  \bibinfo {author} {\bibfnamefont {G.}~\bibnamefont {Chen}},\ }\href {\doibase
  10.1088/0305-4470/34/7/317} {\bibfield  {journal} {\bibinfo  {journal} {J.
  Phys. A}\ }\textbf {\bibinfo {volume} {34}},\ \bibinfo {pages} {1459}
  (\bibinfo {year} {2001})},\ \Eprint {http://arxiv.org/abs/math-ph/0205003}
  {arXiv:math-ph/0205003} \BibitemShut {NoStop}%
\bibitem [{\citenamefont {Guerra}\ \emph {et~al.}(2019)\citenamefont {Guerra},
  \citenamefont {Macedo},\ and\ \citenamefont {Pani}}]{Guerra2019}%
  \BibitemOpen
  \bibfield  {author} {\bibinfo {author} {\bibfnamefont {D.}~\bibnamefont
  {Guerra}}, \bibinfo {author} {\bibfnamefont {C.~F.}\ \bibnamefont {Macedo}},
  \ and\ \bibinfo {author} {\bibfnamefont {P.}~\bibnamefont {Pani}},\ }\href
  {\doibase 10.1088/1475-7516/2019/09/061} {\bibfield  {journal} {\bibinfo
  {journal} {Journal of Cosmology and Astroparticle Physics}\ }\textbf
  {\bibinfo {volume} {2019}},\ \bibinfo {pages} {061} (\bibinfo {year}
  {2019})}\BibitemShut {NoStop}%
\bibitem [{\citenamefont {Lee}\ and\ \citenamefont {Pang}(1989)}]{Lee1989}%
  \BibitemOpen
  \bibfield  {author} {\bibinfo {author} {\bibfnamefont {T.}~\bibnamefont
  {Lee}}\ and\ \bibinfo {author} {\bibfnamefont {Y.}~\bibnamefont {Pang}},\
  }\href {\doibase 10.1016/0550-3213(89)90365-9} {\bibfield  {journal}
  {\bibinfo  {journal} {Nuclear Physics B}\ }\textbf {\bibinfo {volume}
  {315}},\ \bibinfo {pages} {477} (\bibinfo {year} {1989})}\BibitemShut
  {NoStop}%
\bibitem [{\citenamefont {Brito}\ \emph {et~al.}(2016)\citenamefont {Brito},
  \citenamefont {Cardoso}, \citenamefont {Herdeiro},\ and\ \citenamefont
  {Radu}}]{Brito2016}%
  \BibitemOpen
  \bibfield  {author} {\bibinfo {author} {\bibfnamefont {R.}~\bibnamefont
  {Brito}}, \bibinfo {author} {\bibfnamefont {V.}~\bibnamefont {Cardoso}},
  \bibinfo {author} {\bibfnamefont {C.~A.}\ \bibnamefont {Herdeiro}}, \ and\
  \bibinfo {author} {\bibfnamefont {E.}~\bibnamefont {Radu}},\ }\href {\doibase
  10.1016/j.physletb.2015.11.051} {\bibfield  {journal} {\bibinfo  {journal}
  {Physics Letters B}\ }\textbf {\bibinfo {volume} {752}},\ \bibinfo {pages}
  {291} (\bibinfo {year} {2016})}\BibitemShut {NoStop}%
\bibitem [{\citenamefont {Herdeiro}\ and\ \citenamefont
  {Radu}(2015{\natexlab{a}})}]{Herdeiro2015}%
  \BibitemOpen
  \bibfield  {author} {\bibinfo {author} {\bibfnamefont {C.~A.~R.}\
  \bibnamefont {Herdeiro}}\ and\ \bibinfo {author} {\bibfnamefont
  {E.}~\bibnamefont {Radu}},\ }\href {\doibase 10.1142/s0218271815420146}
  {\bibfield  {journal} {\bibinfo  {journal} {International Journal of Modern
  Physics D}\ }\textbf {\bibinfo {volume} {24}},\ \bibinfo {pages} {1542014}
  (\bibinfo {year} {2015}{\natexlab{a}})}\BibitemShut {NoStop}%
\bibitem [{\citenamefont {Henriques}\ \emph {et~al.}(1989)\citenamefont
  {Henriques}, \citenamefont {Liddle},\ and\ \citenamefont
  {Moorhouse}}]{Henriques1989}%
  \BibitemOpen
  \bibfield  {author} {\bibinfo {author} {\bibfnamefont {A.}~\bibnamefont
  {Henriques}}, \bibinfo {author} {\bibfnamefont {A.~R.}\ \bibnamefont
  {Liddle}}, \ and\ \bibinfo {author} {\bibfnamefont {R.}~\bibnamefont
  {Moorhouse}},\ }\href {\doibase 10.1016/0370-2693(89)90623-0} {\bibfield
  {journal} {\bibinfo  {journal} {Physics Letters B}\ }\textbf {\bibinfo
  {volume} {233}},\ \bibinfo {pages} {99} (\bibinfo {year} {1989})}\BibitemShut
  {NoStop}%
\bibitem [{\citenamefont {Bernal}\ \emph {et~al.}(2010)\citenamefont {Bernal},
  \citenamefont {Barranco}, \citenamefont {Alic},\ and\ \citenamefont
  {Palenzuela}}]{Bernal2010}%
  \BibitemOpen
  \bibfield  {author} {\bibinfo {author} {\bibfnamefont {A.}~\bibnamefont
  {Bernal}}, \bibinfo {author} {\bibfnamefont {J.}~\bibnamefont {Barranco}},
  \bibinfo {author} {\bibfnamefont {D.}~\bibnamefont {Alic}}, \ and\ \bibinfo
  {author} {\bibfnamefont {C.}~\bibnamefont {Palenzuela}},\ }\href {\doibase
  10.1103/physrevd.81.044031} {\bibfield  {journal} {\bibinfo  {journal}
  {Physical Review D}\ }\textbf {\bibinfo {volume} {81}},\ \bibinfo {pages}
  {044031} (\bibinfo {year} {2010})}\BibitemShut {NoStop}%
\bibitem [{\citenamefont {Li}\ \emph {et~al.}(2020)\citenamefont {Li},
  \citenamefont {Sun}, \citenamefont {Hu}, \citenamefont {Song},\ and\
  \citenamefont {Wang}}]{Li:2019mlk}%
  \BibitemOpen
  \bibfield  {author} {\bibinfo {author} {\bibfnamefont {H.-B.}\ \bibnamefont
  {Li}}, \bibinfo {author} {\bibfnamefont {S.}~\bibnamefont {Sun}}, \bibinfo
  {author} {\bibfnamefont {T.-T.}\ \bibnamefont {Hu}}, \bibinfo {author}
  {\bibfnamefont {Y.}~\bibnamefont {Song}}, \ and\ \bibinfo {author}
  {\bibfnamefont {Y.-Q.}\ \bibnamefont {Wang}},\ }\href {\doibase
  10.1103/PhysRevD.101.044017} {\bibfield  {journal} {\bibinfo  {journal}
  {Phys. Rev. D}\ }\textbf {\bibinfo {volume} {101}},\ \bibinfo {pages}
  {044017} (\bibinfo {year} {2020})},\ \Eprint
  {http://arxiv.org/abs/1906.00420} {arXiv:1906.00420 [gr-qc]} \BibitemShut
  {NoStop}%
\bibitem [{\citenamefont {Li}\ \emph {et~al.}(2021)\citenamefont {Li},
  \citenamefont {Zeng}, \citenamefont {Song},\ and\ \citenamefont
  {Wang}}]{Li:2020ffy}%
  \BibitemOpen
  \bibfield  {author} {\bibinfo {author} {\bibfnamefont {H.-B.}\ \bibnamefont
  {Li}}, \bibinfo {author} {\bibfnamefont {Y.-B.}\ \bibnamefont {Zeng}},
  \bibinfo {author} {\bibfnamefont {Y.}~\bibnamefont {Song}}, \ and\ \bibinfo
  {author} {\bibfnamefont {Y.-Q.}\ \bibnamefont {Wang}},\ }\href {\doibase
  10.1007/JHEP04(2021)042} {\bibfield  {journal} {\bibinfo  {journal} {JHEP}\
  }\textbf {\bibinfo {volume} {04}},\ \bibinfo {pages} {042} (\bibinfo {year}
  {2021})},\ \Eprint {http://arxiv.org/abs/2006.11281} {arXiv:2006.11281
  [gr-qc]} \BibitemShut {NoStop}%
\bibitem [{\citenamefont {Zeng}\ \emph {et~al.}(2023)\citenamefont {Zeng},
  \citenamefont {Sun}, \citenamefont {Cui}, \citenamefont {Zhang},\ and\
  \citenamefont {Wang}}]{Zeng:2023hvq}%
  \BibitemOpen
  \bibfield  {author} {\bibinfo {author} {\bibfnamefont {Y.-B.}\ \bibnamefont
  {Zeng}}, \bibinfo {author} {\bibfnamefont {S.-X.}\ \bibnamefont {Sun}},
  \bibinfo {author} {\bibfnamefont {S.-Y.}\ \bibnamefont {Cui}}, \bibinfo
  {author} {\bibfnamefont {Y.-P.}\ \bibnamefont {Zhang}}, \ and\ \bibinfo
  {author} {\bibfnamefont {Y.-Q.}\ \bibnamefont {Wang}},\ }\href@noop {} {\
  (\bibinfo {year} {2023})},\ \Eprint {http://arxiv.org/abs/2309.05743}
  {arXiv:2309.05743 [gr-qc]} \BibitemShut {NoStop}%
\bibitem [{\citenamefont {Braaten}\ \emph {et~al.}(2016)\citenamefont
  {Braaten}, \citenamefont {Mohapatra},\ and\ \citenamefont
  {Zhang}}]{Braaten:2015eeu}%
  \BibitemOpen
  \bibfield  {author} {\bibinfo {author} {\bibfnamefont {E.}~\bibnamefont
  {Braaten}}, \bibinfo {author} {\bibfnamefont {A.}~\bibnamefont {Mohapatra}},
  \ and\ \bibinfo {author} {\bibfnamefont {H.}~\bibnamefont {Zhang}},\ }\href
  {\doibase 10.1103/PhysRevLett.117.121801} {\bibfield  {journal} {\bibinfo
  {journal} {Phys. Rev. Lett.}\ }\textbf {\bibinfo {volume} {117}},\ \bibinfo
  {pages} {121801} (\bibinfo {year} {2016})},\ \Eprint
  {http://arxiv.org/abs/1512.00108} {arXiv:1512.00108 [hep-ph]} \BibitemShut
  {NoStop}%
\bibitem [{\citenamefont {Zhang}(2019)}]{Zhang:2018slz}%
  \BibitemOpen
  \bibfield  {author} {\bibinfo {author} {\bibfnamefont {H.}~\bibnamefont
  {Zhang}},\ }\href {\doibase 10.3390/sym12010025} {\bibfield  {journal}
  {\bibinfo  {journal} {Symmetry}\ }\textbf {\bibinfo {volume} {12}},\ \bibinfo
  {pages} {25} (\bibinfo {year} {2019})},\ \Eprint
  {http://arxiv.org/abs/1810.11473} {arXiv:1810.11473 [hep-ph]} \BibitemShut
  {NoStop}%
\bibitem [{\citenamefont {Braaten}\ and\ \citenamefont
  {Zhang}(2019)}]{Braaten:2019knj}%
  \BibitemOpen
  \bibfield  {author} {\bibinfo {author} {\bibfnamefont {E.}~\bibnamefont
  {Braaten}}\ and\ \bibinfo {author} {\bibfnamefont {H.}~\bibnamefont
  {Zhang}},\ }\href {\doibase 10.1103/RevModPhys.91.041002} {\bibfield
  {journal} {\bibinfo  {journal} {Rev. Mod. Phys.}\ }\textbf {\bibinfo {volume}
  {91}},\ \bibinfo {pages} {041002} (\bibinfo {year} {2019})}\BibitemShut
  {NoStop}%
\bibitem [{\citenamefont {Hawley}\ and\ \citenamefont
  {Choptuik}(2003)}]{Hawley2003}%
  \BibitemOpen
  \bibfield  {author} {\bibinfo {author} {\bibfnamefont {S.~H.}\ \bibnamefont
  {Hawley}}\ and\ \bibinfo {author} {\bibfnamefont {M.~W.}\ \bibnamefont
  {Choptuik}},\ }\href {\doibase 10.1103/physrevd.67.024010} {\bibfield
  {journal} {\bibinfo  {journal} {Physical Review D}\ }\textbf {\bibinfo
  {volume} {67}},\ \bibinfo {pages} {024010} (\bibinfo {year}
  {2003})}\BibitemShut {NoStop}%
\bibitem [{\citenamefont {Schunck}\ and\ \citenamefont
  {Mielke}(1996)}]{Schunck1996}%
  \BibitemOpen
  \bibfield  {author} {\bibinfo {author} {\bibfnamefont {F.~E.}\ \bibnamefont
  {Schunck}}\ and\ \bibinfo {author} {\bibfnamefont {E.~W.}\ \bibnamefont
  {Mielke}},\ }\enquote {\bibinfo {title} {Rotating boson stars},}\ in\ \href
  {\doibase 10.1007/978-3-642-95732-1_7} {\emph {\bibinfo {booktitle}
  {Relativity and Scientific Computing}}}\ (\bibinfo  {publisher} {Springer
  Berlin Heidelberg},\ \bibinfo {year} {1996})\ pp.\ \bibinfo {pages}
  {138--151}\BibitemShut {NoStop}%
\bibitem [{\citenamefont {Yoshida}\ and\ \citenamefont
  {Eriguchi}(1997)}]{Yoshida1997}%
  \BibitemOpen
  \bibfield  {author} {\bibinfo {author} {\bibfnamefont {S.}~\bibnamefont
  {Yoshida}}\ and\ \bibinfo {author} {\bibfnamefont {Y.}~\bibnamefont
  {Eriguchi}},\ }\href {\doibase 10.1103/physrevd.56.762} {\bibfield  {journal}
  {\bibinfo  {journal} {Physical Review D}\ }\textbf {\bibinfo {volume} {56}},\
  \bibinfo {pages} {762} (\bibinfo {year} {1997})}\BibitemShut {NoStop}%
\bibitem [{\citenamefont {Volkov}\ and\ \citenamefont
  {Wohnert}(2002)}]{Volkov2002}%
  \BibitemOpen
  \bibfield  {author} {\bibinfo {author} {\bibfnamefont {M.~S.}\ \bibnamefont
  {Volkov}}\ and\ \bibinfo {author} {\bibfnamefont {E.}~\bibnamefont
  {Wohnert}},\ }\href {\doibase 10.1103/PhysRevD.66.085003} {\bibfield
  {journal} {\bibinfo  {journal} {Phys. Rev. D}\ }\textbf {\bibinfo {volume}
  {66}},\ \bibinfo {pages} {085003} (\bibinfo {year} {2002})},\ \Eprint
  {http://arxiv.org/abs/hep-th/0205157} {arXiv:hep-th/0205157} \BibitemShut
  {NoStop}%
\bibitem [{\citenamefont {Kleihaus}\ \emph {et~al.}(2008)\citenamefont
  {Kleihaus}, \citenamefont {Kunz}, \citenamefont {List},\ and\ \citenamefont
  {Schaffer}}]{Kleihaus:2007vk}%
  \BibitemOpen
  \bibfield  {author} {\bibinfo {author} {\bibfnamefont {B.}~\bibnamefont
  {Kleihaus}}, \bibinfo {author} {\bibfnamefont {J.}~\bibnamefont {Kunz}},
  \bibinfo {author} {\bibfnamefont {M.}~\bibnamefont {List}}, \ and\ \bibinfo
  {author} {\bibfnamefont {I.}~\bibnamefont {Schaffer}},\ }\href {\doibase
  10.1103/PhysRevD.77.064025} {\bibfield  {journal} {\bibinfo  {journal} {Phys.
  Rev. D}\ }\textbf {\bibinfo {volume} {77}},\ \bibinfo {pages} {064025}
  (\bibinfo {year} {2008})},\ \Eprint {http://arxiv.org/abs/0712.3742}
  {arXiv:0712.3742 [gr-qc]} \BibitemShut {NoStop}%
\bibitem [{\citenamefont {Wang}\ \emph {et~al.}(2019)\citenamefont {Wang},
  \citenamefont {Liu},\ and\ \citenamefont {Wei}}]{Wang:2018xhw}%
  \BibitemOpen
  \bibfield  {author} {\bibinfo {author} {\bibfnamefont {Y.-Q.}\ \bibnamefont
  {Wang}}, \bibinfo {author} {\bibfnamefont {Y.-X.}\ \bibnamefont {Liu}}, \
  and\ \bibinfo {author} {\bibfnamefont {S.-W.}\ \bibnamefont {Wei}},\ }\href
  {\doibase 10.1103/PhysRevD.99.064036} {\bibfield  {journal} {\bibinfo
  {journal} {Phys. Rev. D}\ }\textbf {\bibinfo {volume} {99}},\ \bibinfo
  {pages} {064036} (\bibinfo {year} {2019})},\ \Eprint
  {http://arxiv.org/abs/1811.08795} {arXiv:1811.08795 [gr-qc]} \BibitemShut
  {NoStop}%
\bibitem [{\citenamefont {Kunz}\ \emph {et~al.}(2019)\citenamefont {Kunz},
  \citenamefont {Perapechka},\ and\ \citenamefont {Shnir}}]{Kunz:2019bhm}%
  \BibitemOpen
  \bibfield  {author} {\bibinfo {author} {\bibfnamefont {J.}~\bibnamefont
  {Kunz}}, \bibinfo {author} {\bibfnamefont {I.}~\bibnamefont {Perapechka}}, \
  and\ \bibinfo {author} {\bibfnamefont {Y.}~\bibnamefont {Shnir}},\ }\href
  {\doibase 10.1103/PhysRevD.100.064032} {\bibfield  {journal} {\bibinfo
  {journal} {Phys. Rev. D}\ }\textbf {\bibinfo {volume} {100}},\ \bibinfo
  {pages} {064032} (\bibinfo {year} {2019})},\ \Eprint
  {http://arxiv.org/abs/1904.07630} {arXiv:1904.07630 [gr-qc]} \BibitemShut
  {NoStop}%
\bibitem [{\citenamefont {Palenzuela}\ \emph {et~al.}(2017)\citenamefont
  {Palenzuela}, \citenamefont {Pani}, \citenamefont {Bezares}, \citenamefont
  {Cardoso}, \citenamefont {Lehner},\ and\ \citenamefont
  {Liebling}}]{Palenzuela2017}%
  \BibitemOpen
  \bibfield  {author} {\bibinfo {author} {\bibfnamefont {C.}~\bibnamefont
  {Palenzuela}}, \bibinfo {author} {\bibfnamefont {P.}~\bibnamefont {Pani}},
  \bibinfo {author} {\bibfnamefont {M.}~\bibnamefont {Bezares}}, \bibinfo
  {author} {\bibfnamefont {V.}~\bibnamefont {Cardoso}}, \bibinfo {author}
  {\bibfnamefont {L.}~\bibnamefont {Lehner}}, \ and\ \bibinfo {author}
  {\bibfnamefont {S.}~\bibnamefont {Liebling}},\ }\href {\doibase
  10.1103/physrevd.96.104058} {\bibfield  {journal} {\bibinfo  {journal} {Phys.
  Rev. D 96, 104058 (2017)}\ }\textbf {\bibinfo {volume} {96}},\ \bibinfo
  {pages} {104058} (\bibinfo {year} {2017})},\ \Eprint
  {http://arxiv.org/abs/1710.09432} {arXiv:1710.09432 [gr-qc]} \BibitemShut
  {NoStop}%
\bibitem [{\citenamefont {Croft}\ \emph {et~al.}(2023)\citenamefont {Croft},
  \citenamefont {Helfer}, \citenamefont {Ge}, \citenamefont {Radia},
  \citenamefont {Evstafyeva}, \citenamefont {Lim}, \citenamefont {Sperhake},\
  and\ \citenamefont {Clough}}]{Croft2023}%
  \BibitemOpen
  \bibfield  {author} {\bibinfo {author} {\bibfnamefont {R.}~\bibnamefont
  {Croft}}, \bibinfo {author} {\bibfnamefont {T.}~\bibnamefont {Helfer}},
  \bibinfo {author} {\bibfnamefont {B.-X.}\ \bibnamefont {Ge}}, \bibinfo
  {author} {\bibfnamefont {M.}~\bibnamefont {Radia}}, \bibinfo {author}
  {\bibfnamefont {T.}~\bibnamefont {Evstafyeva}}, \bibinfo {author}
  {\bibfnamefont {E.~A.}\ \bibnamefont {Lim}}, \bibinfo {author} {\bibfnamefont
  {U.}~\bibnamefont {Sperhake}}, \ and\ \bibinfo {author} {\bibfnamefont
  {K.}~\bibnamefont {Clough}},\ }\href {\doibase 10.1088/1361-6382/acace4}
  {\bibfield  {journal} {\bibinfo  {journal} {Classical and Quantum Gravity}\
  }\textbf {\bibinfo {volume} {40}},\ \bibinfo {pages} {065001} (\bibinfo
  {year} {2023})}\BibitemShut {NoStop}%
\bibitem [{\citenamefont {Dabrowski}\ and\ \citenamefont
  {Schunck}(2000)}]{Dabrowski:1998ac}%
  \BibitemOpen
  \bibfield  {author} {\bibinfo {author} {\bibfnamefont {M.~P.}\ \bibnamefont
  {Dabrowski}}\ and\ \bibinfo {author} {\bibfnamefont {F.~E.}\ \bibnamefont
  {Schunck}},\ }\href {\doibase 10.1086/308805} {\bibfield  {journal} {\bibinfo
   {journal} {Astrophys. J.}\ }\textbf {\bibinfo {volume} {535}},\ \bibinfo
  {pages} {316} (\bibinfo {year} {2000})},\ \Eprint
  {http://arxiv.org/abs/astro-ph/9807039} {arXiv:astro-ph/9807039} \BibitemShut
  {NoStop}%
\bibitem [{\citenamefont {Rosa}\ \emph {et~al.}(2022)\citenamefont {Rosa},
  \citenamefont {Garcia}, \citenamefont {Vincent},\ and\ \citenamefont
  {Cardoso}}]{Rosa:2022toh}%
  \BibitemOpen
  \bibfield  {author} {\bibinfo {author} {\bibfnamefont {J.~a.~L.}\
  \bibnamefont {Rosa}}, \bibinfo {author} {\bibfnamefont {P.}~\bibnamefont
  {Garcia}}, \bibinfo {author} {\bibfnamefont {F.~H.}\ \bibnamefont {Vincent}},
  \ and\ \bibinfo {author} {\bibfnamefont {V.}~\bibnamefont {Cardoso}},\ }\href
  {\doibase 10.1103/PhysRevD.106.044031} {\bibfield  {journal} {\bibinfo
  {journal} {Phys. Rev. D}\ }\textbf {\bibinfo {volume} {106}},\ \bibinfo
  {pages} {044031} (\bibinfo {year} {2022})},\ \Eprint
  {http://arxiv.org/abs/2205.11541} {arXiv:2205.11541 [gr-qc]} \BibitemShut
  {NoStop}%
\bibitem [{\citenamefont {Rosa}\ and\ \citenamefont
  {Rubiera-Garcia}(2022)}]{Rosa:2022tfv}%
  \BibitemOpen
  \bibfield  {author} {\bibinfo {author} {\bibfnamefont {J.~a.~L.}\
  \bibnamefont {Rosa}}\ and\ \bibinfo {author} {\bibfnamefont {D.}~\bibnamefont
  {Rubiera-Garcia}},\ }\href {\doibase 10.1103/PhysRevD.106.084004} {\bibfield
  {journal} {\bibinfo  {journal} {Phys. Rev. D}\ }\textbf {\bibinfo {volume}
  {106}},\ \bibinfo {pages} {084004} (\bibinfo {year} {2022})},\ \Eprint
  {http://arxiv.org/abs/2204.12949} {arXiv:2204.12949 [gr-qc]} \BibitemShut
  {NoStop}%
\bibitem [{\citenamefont {Rosa}\ \emph {et~al.}(2023)\citenamefont {Rosa},
  \citenamefont {Macedo},\ and\ \citenamefont {Rubiera-Garcia}}]{Rosa:2023qcv}%
  \BibitemOpen
  \bibfield  {author} {\bibinfo {author} {\bibfnamefont {J.~a.~L.}\
  \bibnamefont {Rosa}}, \bibinfo {author} {\bibfnamefont {C.~F.~B.}\
  \bibnamefont {Macedo}}, \ and\ \bibinfo {author} {\bibfnamefont
  {D.}~\bibnamefont {Rubiera-Garcia}},\ }\href {\doibase
  10.1103/PhysRevD.108.044021} {\bibfield  {journal} {\bibinfo  {journal}
  {Phys. Rev. D}\ }\textbf {\bibinfo {volume} {108}},\ \bibinfo {pages}
  {044021} (\bibinfo {year} {2023})},\ \Eprint
  {http://arxiv.org/abs/2303.17296} {arXiv:2303.17296 [gr-qc]} \BibitemShut
  {NoStop}%
\bibitem [{\citenamefont {Rosa}\ \emph {et~al.}(2024)\citenamefont {Rosa},
  \citenamefont {Pelle},\ and\ \citenamefont {P\'erez}}]{Rosa:2024eva}%
  \BibitemOpen
  \bibfield  {author} {\bibinfo {author} {\bibfnamefont {J.~a.~L.}\
  \bibnamefont {Rosa}}, \bibinfo {author} {\bibfnamefont {J.}~\bibnamefont
  {Pelle}}, \ and\ \bibinfo {author} {\bibfnamefont {D.}~\bibnamefont
  {P\'erez}},\ }\href {\doibase 10.1103/PhysRevD.110.084068} {\bibfield
  {journal} {\bibinfo  {journal} {Phys. Rev. D}\ }\textbf {\bibinfo {volume}
  {110}},\ \bibinfo {pages} {084068} (\bibinfo {year} {2024})},\ \Eprint
  {http://arxiv.org/abs/2403.11540} {arXiv:2403.11540 [gr-qc]} \BibitemShut
  {NoStop}%
\bibitem [{\citenamefont {Cunha}\ \emph
  {et~al.}(2016{\natexlab{a}})\citenamefont {Cunha}, \citenamefont {Grover},
  \citenamefont {Herdeiro}, \citenamefont {Radu}, \citenamefont {Runarsson},\
  and\ \citenamefont {Wittig}}]{Cunha:2016bjh}%
  \BibitemOpen
  \bibfield  {author} {\bibinfo {author} {\bibfnamefont {P.~V.~P.}\
  \bibnamefont {Cunha}}, \bibinfo {author} {\bibfnamefont {J.}~\bibnamefont
  {Grover}}, \bibinfo {author} {\bibfnamefont {C.}~\bibnamefont {Herdeiro}},
  \bibinfo {author} {\bibfnamefont {E.}~\bibnamefont {Radu}}, \bibinfo {author}
  {\bibfnamefont {H.}~\bibnamefont {Runarsson}}, \ and\ \bibinfo {author}
  {\bibfnamefont {A.}~\bibnamefont {Wittig}},\ }\href {\doibase
  10.1103/PhysRevD.94.104023} {\bibfield  {journal} {\bibinfo  {journal} {Phys.
  Rev. D}\ }\textbf {\bibinfo {volume} {94}},\ \bibinfo {pages} {104023}
  (\bibinfo {year} {2016}{\natexlab{a}})},\ \Eprint
  {http://arxiv.org/abs/1609.01340} {arXiv:1609.01340 [gr-qc]} \BibitemShut
  {NoStop}%
\bibitem [{\citenamefont {Chen}\ \emph {et~al.}(2023)\citenamefont {Chen},
  \citenamefont {Jing}, \citenamefont {Qian},\ and\ \citenamefont
  {Wang}}]{Chen:2022scf}%
  \BibitemOpen
  \bibfield  {author} {\bibinfo {author} {\bibfnamefont {S.}~\bibnamefont
  {Chen}}, \bibinfo {author} {\bibfnamefont {J.}~\bibnamefont {Jing}}, \bibinfo
  {author} {\bibfnamefont {W.-L.}\ \bibnamefont {Qian}}, \ and\ \bibinfo
  {author} {\bibfnamefont {B.}~\bibnamefont {Wang}},\ }\href {\doibase
  10.1007/s11433-022-2059-5} {\bibfield  {journal} {\bibinfo  {journal} {Sci.
  China Phys. Mech. Astron.}\ }\textbf {\bibinfo {volume} {66}},\ \bibinfo
  {pages} {260401} (\bibinfo {year} {2023})},\ \Eprint
  {http://arxiv.org/abs/2301.00113} {arXiv:2301.00113 [astro-ph.HE]}
  \BibitemShut {NoStop}%
\bibitem [{\citenamefont {Cunha}\ \emph {et~al.}(2017)\citenamefont {Cunha},
  \citenamefont {Font}, \citenamefont {Herdeiro}, \citenamefont {Radu},
  \citenamefont {Sanchis-Gual},\ and\ \citenamefont {Zilhão}}]{Cunha2017}%
  \BibitemOpen
  \bibfield  {author} {\bibinfo {author} {\bibfnamefont {P.~V.~P.}\
  \bibnamefont {Cunha}}, \bibinfo {author} {\bibfnamefont {J.~A.}\ \bibnamefont
  {Font}}, \bibinfo {author} {\bibfnamefont {C.}~\bibnamefont {Herdeiro}},
  \bibinfo {author} {\bibfnamefont {E.}~\bibnamefont {Radu}}, \bibinfo {author}
  {\bibfnamefont {N.}~\bibnamefont {Sanchis-Gual}}, \ and\ \bibinfo {author}
  {\bibfnamefont {M.}~\bibnamefont {Zilhão}},\ }\href {\doibase
  10.1103/physrevd.96.104040} {\bibfield  {journal} {\bibinfo  {journal} {Phys.
  Rev. D 96, 104040 (2017)}\ }\textbf {\bibinfo {volume} {96}},\ \bibinfo
  {pages} {104040} (\bibinfo {year} {2017})},\ \Eprint
  {http://arxiv.org/abs/1709.06118} {1709.06118 [gr-qc]} \BibitemShut {NoStop}%
\bibitem [{\citenamefont {Hongsheng}\ and\ \citenamefont
  {Xilong}(2021)}]{Hongsheng:2018ibg}%
  \BibitemOpen
  \bibfield  {author} {\bibinfo {author} {\bibfnamefont {Z.}~\bibnamefont
  {Hongsheng}}\ and\ \bibinfo {author} {\bibfnamefont {F.}~\bibnamefont
  {Xilong}},\ }\href {\doibase 10.1007/s11433-021-1764-y} {\bibfield  {journal}
  {\bibinfo  {journal} {Sci. China Phys. Mech. Astron.}\ }\textbf {\bibinfo
  {volume} {64}},\ \bibinfo {pages} {120462} (\bibinfo {year} {2021})},\
  \Eprint {http://arxiv.org/abs/1809.06511} {arXiv:1809.06511 [gr-qc]}
  \BibitemShut {NoStop}%
\bibitem [{\citenamefont {Herdeiro}\ and\ \citenamefont
  {Radu}(2023)}]{Herdeiro:2023roz}%
  \BibitemOpen
  \bibfield  {author} {\bibinfo {author} {\bibfnamefont {C.~A.~R.}\
  \bibnamefont {Herdeiro}}\ and\ \bibinfo {author} {\bibfnamefont
  {E.}~\bibnamefont {Radu}},\ }\href {\doibase 10.1103/PhysRevLett.131.121401}
  {\bibfield  {journal} {\bibinfo  {journal} {Phys. Rev. Lett.}\ }\textbf
  {\bibinfo {volume} {131}},\ \bibinfo {pages} {121401} (\bibinfo {year}
  {2023})},\ \Eprint {http://arxiv.org/abs/2305.15467} {arXiv:2305.15467
  [gr-qc]} \BibitemShut {NoStop}%
\bibitem [{\citenamefont {Herdeiro}\ and\ \citenamefont
  {Radu}(2014)}]{Herdeiro:2014goa}%
  \BibitemOpen
  \bibfield  {author} {\bibinfo {author} {\bibfnamefont {C.~A.~R.}\
  \bibnamefont {Herdeiro}}\ and\ \bibinfo {author} {\bibfnamefont
  {E.}~\bibnamefont {Radu}},\ }\href {\doibase 10.1103/PhysRevLett.112.221101}
  {\bibfield  {journal} {\bibinfo  {journal} {Phys. Rev. Lett.}\ }\textbf
  {\bibinfo {volume} {112}},\ \bibinfo {pages} {221101} (\bibinfo {year}
  {2014})},\ \Eprint {http://arxiv.org/abs/1403.2757} {arXiv:1403.2757 [gr-qc]}
  \BibitemShut {NoStop}%
\bibitem [{\citenamefont {Herdeiro}\ and\ \citenamefont
  {Radu}(2015{\natexlab{b}})}]{Herdeiro:2015gia}%
  \BibitemOpen
  \bibfield  {author} {\bibinfo {author} {\bibfnamefont {C.}~\bibnamefont
  {Herdeiro}}\ and\ \bibinfo {author} {\bibfnamefont {E.}~\bibnamefont
  {Radu}},\ }\href {\doibase 10.1088/0264-9381/32/14/144001} {\bibfield
  {journal} {\bibinfo  {journal} {Class. Quant. Grav.}\ }\textbf {\bibinfo
  {volume} {32}},\ \bibinfo {pages} {144001} (\bibinfo {year}
  {2015}{\natexlab{b}})},\ \Eprint {http://arxiv.org/abs/1501.04319}
  {arXiv:1501.04319 [gr-qc]} \BibitemShut {NoStop}%
\bibitem [{\citenamefont {Fernandes}\ and\ \citenamefont
  {Mulryne}(2023)}]{Fernandes:2022gde}%
  \BibitemOpen
  \bibfield  {author} {\bibinfo {author} {\bibfnamefont {P.~G.~S.}\
  \bibnamefont {Fernandes}}\ and\ \bibinfo {author} {\bibfnamefont {D.~J.}\
  \bibnamefont {Mulryne}},\ }\href {\doibase 10.1088/1361-6382/ace232}
  {\bibfield  {journal} {\bibinfo  {journal} {Class. Quant. Grav.}\ }\textbf
  {\bibinfo {volume} {40}},\ \bibinfo {pages} {165001} (\bibinfo {year}
  {2023})},\ \Eprint {http://arxiv.org/abs/2212.07293} {arXiv:2212.07293
  [gr-qc]} \BibitemShut {NoStop}%
\bibitem [{\citenamefont {Huang}\ \emph {et~al.}(2018)\citenamefont {Huang},
  \citenamefont {Dong},\ and\ \citenamefont {Liu}}]{Huang2018}%
  \BibitemOpen
  \bibfield  {author} {\bibinfo {author} {\bibfnamefont {Y.}~\bibnamefont
  {Huang}}, \bibinfo {author} {\bibfnamefont {Y.-P.}\ \bibnamefont {Dong}}, \
  and\ \bibinfo {author} {\bibfnamefont {D.-J.}\ \bibnamefont {Liu}},\ }\href
  {\doibase 10.1142/S0218271818501146} {\bibfield  {journal} {\bibinfo
  {journal} {Int. J. Mod. Phys. D}\ }\textbf {\bibinfo {volume} {27}},\
  \bibinfo {pages} {1850114} (\bibinfo {year} {2018})},\ \Eprint
  {http://arxiv.org/abs/1807.06268} {arXiv:1807.06268 [gr-qc]} \BibitemShut
  {NoStop}%
\bibitem [{\citenamefont {Cunha}\ \emph {et~al.}(2015)\citenamefont {Cunha},
  \citenamefont {Herdeiro}, \citenamefont {Radu},\ and\ \citenamefont
  {Runarsson}}]{Cunha:2015yba}%
  \BibitemOpen
  \bibfield  {author} {\bibinfo {author} {\bibfnamefont {P.~V.~P.}\
  \bibnamefont {Cunha}}, \bibinfo {author} {\bibfnamefont {C.~A.~R.}\
  \bibnamefont {Herdeiro}}, \bibinfo {author} {\bibfnamefont {E.}~\bibnamefont
  {Radu}}, \ and\ \bibinfo {author} {\bibfnamefont {H.~F.}\ \bibnamefont
  {Runarsson}},\ }\href {\doibase 10.1103/PhysRevLett.115.211102} {\bibfield
  {journal} {\bibinfo  {journal} {Phys. Rev. Lett.}\ }\textbf {\bibinfo
  {volume} {115}},\ \bibinfo {pages} {211102} (\bibinfo {year} {2015})},\
  \Eprint {http://arxiv.org/abs/1509.00021} {arXiv:1509.00021 [gr-qc]}
  \BibitemShut {NoStop}%
\bibitem [{\citenamefont {Cunha}\ \emph
  {et~al.}(2016{\natexlab{b}})\citenamefont {Cunha}, \citenamefont {Herdeiro},
  \citenamefont {Radu},\ and\ \citenamefont {Runarsson}}]{Cunha:2016bpi}%
  \BibitemOpen
  \bibfield  {author} {\bibinfo {author} {\bibfnamefont {P.~V.~P.}\
  \bibnamefont {Cunha}}, \bibinfo {author} {\bibfnamefont {C.~A.~R.}\
  \bibnamefont {Herdeiro}}, \bibinfo {author} {\bibfnamefont {E.}~\bibnamefont
  {Radu}}, \ and\ \bibinfo {author} {\bibfnamefont {H.~F.}\ \bibnamefont
  {Runarsson}},\ }\href {\doibase 10.1142/S0218271816410212} {\bibfield
  {journal} {\bibinfo  {journal} {Int. J. Mod. Phys. D}\ }\textbf {\bibinfo
  {volume} {25}},\ \bibinfo {pages} {1641021} (\bibinfo {year}
  {2016}{\natexlab{b}})},\ \Eprint {http://arxiv.org/abs/1605.08293}
  {arXiv:1605.08293 [gr-qc]} \BibitemShut {NoStop}%
\bibitem [{\citenamefont {Dolence}\ \emph {et~al.}(2009)\citenamefont
  {Dolence}, \citenamefont {Gammie}, \citenamefont {Moscibrodzka},\ and\
  \citenamefont {Leung}}]{Dolence:2009zz}%
  \BibitemOpen
  \bibfield  {author} {\bibinfo {author} {\bibfnamefont {J.~C.}\ \bibnamefont
  {Dolence}}, \bibinfo {author} {\bibfnamefont {C.~F.}\ \bibnamefont {Gammie}},
  \bibinfo {author} {\bibfnamefont {M.}~\bibnamefont {Moscibrodzka}}, \ and\
  \bibinfo {author} {\bibfnamefont {P.~K.}\ \bibnamefont {Leung}},\ }\href
  {\doibase 10.1088/0067-0049/184/2/387} {\bibfield  {journal} {\bibinfo
  {journal} {Astrophys. J. Suppl.}\ }\textbf {\bibinfo {volume} {184}},\
  \bibinfo {pages} {387} (\bibinfo {year} {2009})},\ \Eprint
  {http://arxiv.org/abs/0909.0708} {arXiv:0909.0708 [astro-ph.HE]} \BibitemShut
  {NoStop}%
\bibitem [{\citenamefont {Bronzwaer}\ \emph {et~al.}(2018)\citenamefont
  {Bronzwaer}, \citenamefont {Davelaar}, \citenamefont {Younsi}, \citenamefont
  {Mo\'scibrodzka}, \citenamefont {Falcke}, \citenamefont {Kramer},\ and\
  \citenamefont {Rezzolla}}]{Bronzwaer:2018lde}%
  \BibitemOpen
  \bibfield  {author} {\bibinfo {author} {\bibfnamefont {T.}~\bibnamefont
  {Bronzwaer}}, \bibinfo {author} {\bibfnamefont {J.}~\bibnamefont {Davelaar}},
  \bibinfo {author} {\bibfnamefont {Z.}~\bibnamefont {Younsi}}, \bibinfo
  {author} {\bibfnamefont {M.}~\bibnamefont {Mo\'scibrodzka}}, \bibinfo
  {author} {\bibfnamefont {H.}~\bibnamefont {Falcke}}, \bibinfo {author}
  {\bibfnamefont {M.}~\bibnamefont {Kramer}}, \ and\ \bibinfo {author}
  {\bibfnamefont {L.}~\bibnamefont {Rezzolla}},\ }\href {\doibase
  10.1051/0004-6361/201732149} {\bibfield  {journal} {\bibinfo  {journal}
  {Astron. Astrophys.}\ }\textbf {\bibinfo {volume} {613}},\ \bibinfo {pages}
  {A2} (\bibinfo {year} {2018})},\ \Eprint {http://arxiv.org/abs/1801.10452}
  {arXiv:1801.10452 [astro-ph.HE]} \BibitemShut {NoStop}%
\bibitem [{\citenamefont {Bohn}\ \emph {et~al.}(2015)\citenamefont {Bohn},
  \citenamefont {Throwe}, \citenamefont {H\'ebert}, \citenamefont {Henriksson},
  \citenamefont {Bunandar}, \citenamefont {Scheel},\ and\ \citenamefont
  {Taylor}}]{Bohn2015}%
  \BibitemOpen
  \bibfield  {author} {\bibinfo {author} {\bibfnamefont {A.}~\bibnamefont
  {Bohn}}, \bibinfo {author} {\bibfnamefont {W.}~\bibnamefont {Throwe}},
  \bibinfo {author} {\bibfnamefont {F.}~\bibnamefont {H\'ebert}}, \bibinfo
  {author} {\bibfnamefont {K.}~\bibnamefont {Henriksson}}, \bibinfo {author}
  {\bibfnamefont {D.}~\bibnamefont {Bunandar}}, \bibinfo {author}
  {\bibfnamefont {M.~A.}\ \bibnamefont {Scheel}}, \ and\ \bibinfo {author}
  {\bibfnamefont {N.~W.}\ \bibnamefont {Taylor}},\ }\href {\doibase
  10.1088/0264-9381/32/6/065002} {\bibfield  {journal} {\bibinfo  {journal}
  {Class. Quant. Grav.}\ }\textbf {\bibinfo {volume} {32}},\ \bibinfo {pages}
  {065002} (\bibinfo {year} {2015})},\ \Eprint {http://arxiv.org/abs/1410.7775}
  {arXiv:1410.7775 [gr-qc]} \BibitemShut {NoStop}%
\bibitem [{\citenamefont {Chen}\ and\ \citenamefont
  {Jing}(2024)}]{Chen:2023wzv}%
  \BibitemOpen
  \bibfield  {author} {\bibinfo {author} {\bibfnamefont {S.}~\bibnamefont
  {Chen}}\ and\ \bibinfo {author} {\bibfnamefont {J.}~\bibnamefont {Jing}},\
  }\href {\doibase 10.1088/1475-7516/2024/05/023} {\bibfield  {journal}
  {\bibinfo  {journal} {JCAP}\ }\textbf {\bibinfo {volume} {05}},\ \bibinfo
  {pages} {023} (\bibinfo {year} {2024})},\ \Eprint
  {http://arxiv.org/abs/2310.06490} {arXiv:2310.06490 [gr-qc]} \BibitemShut
  {NoStop}%
\bibitem [{\citenamefont {Sun}\ \emph {et~al.}(2023)\citenamefont {Sun},
  \citenamefont {Liu}, \citenamefont {Qian}, \citenamefont {Chen},\ and\
  \citenamefont {Yue}}]{Sun:2023syd}%
  \BibitemOpen
  \bibfield  {author} {\bibinfo {author} {\bibfnamefont {J.}~\bibnamefont
  {Sun}}, \bibinfo {author} {\bibfnamefont {Y.}~\bibnamefont {Liu}}, \bibinfo
  {author} {\bibfnamefont {W.-L.}\ \bibnamefont {Qian}}, \bibinfo {author}
  {\bibfnamefont {S.}~\bibnamefont {Chen}}, \ and\ \bibinfo {author}
  {\bibfnamefont {R.}~\bibnamefont {Yue}},\ }\href {\doibase
  10.1088/1674-1137/aca4bc} {\bibfield  {journal} {\bibinfo  {journal} {Chin.
  Phys. C}\ }\textbf {\bibinfo {volume} {47}},\ \bibinfo {pages} {025104}
  (\bibinfo {year} {2023})}\BibitemShut {NoStop}%
\bibitem [{\citenamefont {Long}\ \emph {et~al.}(2020)\citenamefont {Long},
  \citenamefont {Chen}, \citenamefont {Wang},\ and\ \citenamefont
  {Jing}}]{Long:2020wqj}%
  \BibitemOpen
  \bibfield  {author} {\bibinfo {author} {\bibfnamefont {F.}~\bibnamefont
  {Long}}, \bibinfo {author} {\bibfnamefont {S.}~\bibnamefont {Chen}}, \bibinfo
  {author} {\bibfnamefont {M.}~\bibnamefont {Wang}}, \ and\ \bibinfo {author}
  {\bibfnamefont {J.}~\bibnamefont {Jing}},\ }\href {\doibase
  10.1140/epjc/s10052-020-08744-8} {\bibfield  {journal} {\bibinfo  {journal}
  {Eur. Phys. J. C}\ }\textbf {\bibinfo {volume} {80}},\ \bibinfo {pages}
  {1180} (\bibinfo {year} {2020})},\ \Eprint {http://arxiv.org/abs/2009.07508}
  {arXiv:2009.07508 [gr-qc]} \BibitemShut {NoStop}%
\bibitem [{\citenamefont {Wang}\ \emph
  {et~al.}(2018{\natexlab{a}})\citenamefont {Wang}, \citenamefont {Chen},\ and\
  \citenamefont {Jing}}]{Wang:2018eui}%
  \BibitemOpen
  \bibfield  {author} {\bibinfo {author} {\bibfnamefont {M.}~\bibnamefont
  {Wang}}, \bibinfo {author} {\bibfnamefont {S.}~\bibnamefont {Chen}}, \ and\
  \bibinfo {author} {\bibfnamefont {J.}~\bibnamefont {Jing}},\ }\href {\doibase
  10.1103/PhysRevD.98.104040} {\bibfield  {journal} {\bibinfo  {journal} {Phys.
  Rev. D}\ }\textbf {\bibinfo {volume} {98}},\ \bibinfo {pages} {104040}
  (\bibinfo {year} {2018}{\natexlab{a}})},\ \Eprint
  {http://arxiv.org/abs/1801.02118} {arXiv:1801.02118 [gr-qc]} \BibitemShut
  {NoStop}%
\bibitem [{\citenamefont {Wang}\ \emph
  {et~al.}(2018{\natexlab{b}})\citenamefont {Wang}, \citenamefont {Chen},\ and\
  \citenamefont {Jing}}]{Wang:2017qhh}%
  \BibitemOpen
  \bibfield  {author} {\bibinfo {author} {\bibfnamefont {M.}~\bibnamefont
  {Wang}}, \bibinfo {author} {\bibfnamefont {S.}~\bibnamefont {Chen}}, \ and\
  \bibinfo {author} {\bibfnamefont {J.}~\bibnamefont {Jing}},\ }\href {\doibase
  10.1103/PhysRevD.97.064029} {\bibfield  {journal} {\bibinfo  {journal} {Phys.
  Rev. D}\ }\textbf {\bibinfo {volume} {97}},\ \bibinfo {pages} {064029}
  (\bibinfo {year} {2018}{\natexlab{b}})},\ \Eprint
  {http://arxiv.org/abs/1710.07172} {arXiv:1710.07172 [gr-qc]} \BibitemShut
  {NoStop}%
\bibitem [{\citenamefont {Kuang}\ \emph {et~al.}(2024)\citenamefont {Kuang},
  \citenamefont {Meng}, \citenamefont {Papantonopoulos},\ and\ \citenamefont
  {Wang}}]{Kuang:2024ugn}%
  \BibitemOpen
  \bibfield  {author} {\bibinfo {author} {\bibfnamefont {X.-M.}\ \bibnamefont
  {Kuang}}, \bibinfo {author} {\bibfnamefont {Y.}~\bibnamefont {Meng}},
  \bibinfo {author} {\bibfnamefont {E.}~\bibnamefont {Papantonopoulos}}, \ and\
  \bibinfo {author} {\bibfnamefont {X.-J.}\ \bibnamefont {Wang}},\ }\href
  {\doibase 10.1103/PhysRevD.110.L061503} {\bibfield  {journal} {\bibinfo
  {journal} {Phys. Rev. D}\ }\textbf {\bibinfo {volume} {110}},\ \bibinfo
  {pages} {L061503} (\bibinfo {year} {2024})},\ \Eprint
  {http://arxiv.org/abs/2406.11932} {arXiv:2406.11932 [gr-qc]} \BibitemShut
  {NoStop}%
\bibitem [{\citenamefont {Hu}\ \emph {et~al.}(2021)\citenamefont {Hu},
  \citenamefont {Zhong}, \citenamefont {Li}, \citenamefont {Guo},\ and\
  \citenamefont {Chen}}]{Hu:2020usx}%
  \BibitemOpen
  \bibfield  {author} {\bibinfo {author} {\bibfnamefont {Z.}~\bibnamefont
  {Hu}}, \bibinfo {author} {\bibfnamefont {Z.}~\bibnamefont {Zhong}}, \bibinfo
  {author} {\bibfnamefont {P.-C.}\ \bibnamefont {Li}}, \bibinfo {author}
  {\bibfnamefont {M.}~\bibnamefont {Guo}}, \ and\ \bibinfo {author}
  {\bibfnamefont {B.}~\bibnamefont {Chen}},\ }\href {\doibase
  10.1103/PhysRevD.103.044057} {\bibfield  {journal} {\bibinfo  {journal}
  {Phys. Rev. D}\ }\textbf {\bibinfo {volume} {103}},\ \bibinfo {pages}
  {044057} (\bibinfo {year} {2021})},\ \Eprint
  {http://arxiv.org/abs/2012.07022} {arXiv:2012.07022 [gr-qc]} \BibitemShut
  {NoStop}%
\bibitem [{\citenamefont {Zhang}\ \emph {et~al.}(2023)\citenamefont {Zhang},
  \citenamefont {Yan}, \citenamefont {Guo},\ and\ \citenamefont
  {Chen}}]{Zhang:2022osx}%
  \BibitemOpen
  \bibfield  {author} {\bibinfo {author} {\bibfnamefont {Z.}~\bibnamefont
  {Zhang}}, \bibinfo {author} {\bibfnamefont {H.}~\bibnamefont {Yan}}, \bibinfo
  {author} {\bibfnamefont {M.}~\bibnamefont {Guo}}, \ and\ \bibinfo {author}
  {\bibfnamefont {B.}~\bibnamefont {Chen}},\ }\href {\doibase
  10.1103/PhysRevD.107.024027} {\bibfield  {journal} {\bibinfo  {journal}
  {Phys. Rev. D}\ }\textbf {\bibinfo {volume} {107}},\ \bibinfo {pages}
  {024027} (\bibinfo {year} {2023})},\ \Eprint
  {http://arxiv.org/abs/2206.04430} {arXiv:2206.04430 [gr-qc]} \BibitemShut
  {NoStop}%
\bibitem [{\citenamefont {Zhong}\ \emph {et~al.}(2021)\citenamefont {Zhong},
  \citenamefont {Hu}, \citenamefont {Yan}, \citenamefont {Guo},\ and\
  \citenamefont {Chen}}]{Zhong:2021mty}%
  \BibitemOpen
  \bibfield  {author} {\bibinfo {author} {\bibfnamefont {Z.}~\bibnamefont
  {Zhong}}, \bibinfo {author} {\bibfnamefont {Z.}~\bibnamefont {Hu}}, \bibinfo
  {author} {\bibfnamefont {H.}~\bibnamefont {Yan}}, \bibinfo {author}
  {\bibfnamefont {M.}~\bibnamefont {Guo}}, \ and\ \bibinfo {author}
  {\bibfnamefont {B.}~\bibnamefont {Chen}},\ }\href {\doibase
  10.1103/PhysRevD.104.104028} {\bibfield  {journal} {\bibinfo  {journal}
  {Phys. Rev. D}\ }\textbf {\bibinfo {volume} {104}},\ \bibinfo {pages}
  {104028} (\bibinfo {year} {2021})},\ \Eprint
  {http://arxiv.org/abs/2108.06140} {arXiv:2108.06140 [gr-qc]} \BibitemShut
  {NoStop}%
\bibitem [{\citenamefont {Yang}(2024)}]{Yang:2024ulu}%
  \BibitemOpen
  \bibfield  {author} {\bibinfo {author} {\bibfnamefont {X.}~\bibnamefont
  {Yang}},\ }\href {\doibase 10.1016/j.dark.2024.101467} {\bibfield  {journal}
  {\bibinfo  {journal} {Phys. Dark Univ.}\ }\textbf {\bibinfo {volume} {44}},\
  \bibinfo {pages} {101467} (\bibinfo {year} {2024})}\BibitemShut {NoStop}%
\bibitem [{\citenamefont {Liu}\ \emph {et~al.}(2024{\natexlab{a}})\citenamefont
  {Liu}, \citenamefont {Wu},\ and\ \citenamefont {Wang}}]{Liu:2024lve}%
  \BibitemOpen
  \bibfield  {author} {\bibinfo {author} {\bibfnamefont {W.}~\bibnamefont
  {Liu}}, \bibinfo {author} {\bibfnamefont {D.}~\bibnamefont {Wu}}, \ and\
  \bibinfo {author} {\bibfnamefont {J.}~\bibnamefont {Wang}},\ }\href@noop {}
  {\  (\bibinfo {year} {2024}{\natexlab{a}})},\ \Eprint
  {http://arxiv.org/abs/2407.07416} {arXiv:2407.07416 [gr-qc]} \BibitemShut
  {NoStop}%
\bibitem [{\citenamefont {Liu}\ \emph {et~al.}(2024{\natexlab{b}})\citenamefont
  {Liu}, \citenamefont {Wu}, \citenamefont {Fang}, \citenamefont {Jing},\ and\
  \citenamefont {Wang}}]{Liu:2024lbi}%
  \BibitemOpen
  \bibfield  {author} {\bibinfo {author} {\bibfnamefont {W.}~\bibnamefont
  {Liu}}, \bibinfo {author} {\bibfnamefont {D.}~\bibnamefont {Wu}}, \bibinfo
  {author} {\bibfnamefont {X.}~\bibnamefont {Fang}}, \bibinfo {author}
  {\bibfnamefont {J.}~\bibnamefont {Jing}}, \ and\ \bibinfo {author}
  {\bibfnamefont {J.}~\bibnamefont {Wang}},\ }\href {\doibase
  10.1088/1475-7516/2024/08/035} {\bibfield  {journal} {\bibinfo  {journal}
  {JCAP}\ }\textbf {\bibinfo {volume} {08}},\ \bibinfo {pages} {035} (\bibinfo
  {year} {2024}{\natexlab{b}})},\ \Eprint {http://arxiv.org/abs/2406.00579}
  {arXiv:2406.00579 [gr-qc]} \BibitemShut {NoStop}%
\bibitem [{\citenamefont {Liu}\ \emph {et~al.}(2024{\natexlab{c}})\citenamefont
  {Liu}, \citenamefont {Wu},\ and\ \citenamefont {Wang}}]{Liu:2024soc}%
  \BibitemOpen
  \bibfield  {author} {\bibinfo {author} {\bibfnamefont {W.}~\bibnamefont
  {Liu}}, \bibinfo {author} {\bibfnamefont {D.}~\bibnamefont {Wu}}, \ and\
  \bibinfo {author} {\bibfnamefont {J.}~\bibnamefont {Wang}},\ }\href {\doibase
  10.1016/j.physletb.2024.139052} {\bibfield  {journal} {\bibinfo  {journal}
  {Phys. Lett. B}\ }\textbf {\bibinfo {volume} {858}},\ \bibinfo {pages}
  {139052} (\bibinfo {year} {2024}{\natexlab{c}})},\ \Eprint
  {http://arxiv.org/abs/2408.05569} {arXiv:2408.05569 [gr-qc]} \BibitemShut
  {NoStop}%
\bibitem [{\citenamefont {Yuan}\ \emph {et~al.}(2024)\citenamefont {Yuan},
  \citenamefont {Luo}, \citenamefont {Hu}, \citenamefont {Zhang},\ and\
  \citenamefont {Chen}}]{Yuan:2024ltr}%
  \BibitemOpen
  \bibfield  {author} {\bibinfo {author} {\bibfnamefont {S.}~\bibnamefont
  {Yuan}}, \bibinfo {author} {\bibfnamefont {C.}~\bibnamefont {Luo}}, \bibinfo
  {author} {\bibfnamefont {Z.}~\bibnamefont {Hu}}, \bibinfo {author}
  {\bibfnamefont {Z.}~\bibnamefont {Zhang}}, \ and\ \bibinfo {author}
  {\bibfnamefont {B.}~\bibnamefont {Chen}},\ }\href@noop {} {\  (\bibinfo
  {year} {2024})},\ \Eprint {http://arxiv.org/abs/2403.06886} {arXiv:2403.06886
  [gr-qc]} \BibitemShut {NoStop}%
\bibitem [{\citenamefont {He}\ \emph {et~al.}(2022)\citenamefont {He},
  \citenamefont {Tao}, \citenamefont {Wang}, \citenamefont {Xue},\ and\
  \citenamefont {Zhang}}]{He:2022opa}%
  \BibitemOpen
  \bibfield  {author} {\bibinfo {author} {\bibfnamefont {A.}~\bibnamefont
  {He}}, \bibinfo {author} {\bibfnamefont {J.}~\bibnamefont {Tao}}, \bibinfo
  {author} {\bibfnamefont {P.}~\bibnamefont {Wang}}, \bibinfo {author}
  {\bibfnamefont {Y.}~\bibnamefont {Xue}}, \ and\ \bibinfo {author}
  {\bibfnamefont {L.}~\bibnamefont {Zhang}},\ }\href {\doibase
  10.1140/epjc/s10052-022-10637-x} {\bibfield  {journal} {\bibinfo  {journal}
  {Eur. Phys. J. C}\ }\textbf {\bibinfo {volume} {82}},\ \bibinfo {pages} {683}
  (\bibinfo {year} {2022})},\ \Eprint {http://arxiv.org/abs/2205.12779}
  {arXiv:2205.12779 [gr-qc]} \BibitemShut {NoStop}%
\bibitem [{\citenamefont {Guo}\ \emph {et~al.}(2022)\citenamefont {Guo},
  \citenamefont {Jiang}, \citenamefont {Wang},\ and\ \citenamefont
  {Wu}}]{Guo:2022muy}%
  \BibitemOpen
  \bibfield  {author} {\bibinfo {author} {\bibfnamefont {G.}~\bibnamefont
  {Guo}}, \bibinfo {author} {\bibfnamefont {X.}~\bibnamefont {Jiang}}, \bibinfo
  {author} {\bibfnamefont {P.}~\bibnamefont {Wang}}, \ and\ \bibinfo {author}
  {\bibfnamefont {H.}~\bibnamefont {Wu}},\ }\href {\doibase
  10.1103/PhysRevD.105.124064} {\bibfield  {journal} {\bibinfo  {journal}
  {Phys. Rev. D}\ }\textbf {\bibinfo {volume} {105}},\ \bibinfo {pages}
  {124064} (\bibinfo {year} {2022})},\ \Eprint
  {http://arxiv.org/abs/2204.13948} {arXiv:2204.13948 [gr-qc]} \BibitemShut
  {NoStop}%
\bibitem [{\citenamefont {Ghosh}\ and\ \citenamefont
  {Sarkar}(2021)}]{Ghosh:2021txu}%
  \BibitemOpen
  \bibfield  {author} {\bibinfo {author} {\bibfnamefont {R.}~\bibnamefont
  {Ghosh}}\ and\ \bibinfo {author} {\bibfnamefont {S.}~\bibnamefont {Sarkar}},\
  }\href {\doibase 10.1103/PhysRevD.104.044019} {\bibfield  {journal} {\bibinfo
   {journal} {Phys. Rev. D}\ }\textbf {\bibinfo {volume} {104}},\ \bibinfo
  {pages} {044019} (\bibinfo {year} {2021})},\ \Eprint
  {http://arxiv.org/abs/2107.07370} {arXiv:2107.07370 [gr-qc]} \BibitemShut
  {NoStop}%
\bibitem [{\citenamefont {Meng}\ \emph {et~al.}(2023)\citenamefont {Meng},
  \citenamefont {Fan}, \citenamefont {Li}, \citenamefont {Han},\ and\
  \citenamefont {Zhang}}]{Meng:2023uws}%
  \BibitemOpen
  \bibfield  {author} {\bibinfo {author} {\bibfnamefont {K.}~\bibnamefont
  {Meng}}, \bibinfo {author} {\bibfnamefont {X.-L.}\ \bibnamefont {Fan}},
  \bibinfo {author} {\bibfnamefont {S.}~\bibnamefont {Li}}, \bibinfo {author}
  {\bibfnamefont {W.-B.}\ \bibnamefont {Han}}, \ and\ \bibinfo {author}
  {\bibfnamefont {H.}~\bibnamefont {Zhang}},\ }\href {\doibase
  10.1007/JHEP11(2023)141} {\bibfield  {journal} {\bibinfo  {journal} {JHEP}\
  }\textbf {\bibinfo {volume} {11}},\ \bibinfo {pages} {141} (\bibinfo {year}
  {2023})},\ \Eprint {http://arxiv.org/abs/2307.08953} {arXiv:2307.08953
  [gr-qc]} \BibitemShut {NoStop}%
\bibitem [{\citenamefont {Wang}\ \emph {et~al.}(2024)\citenamefont {Wang},
  \citenamefont {Feng},\ and\ \citenamefont {Wang}}]{Wang:2023fge}%
  \BibitemOpen
  \bibfield  {author} {\bibinfo {author} {\bibfnamefont {K.}~\bibnamefont
  {Wang}}, \bibinfo {author} {\bibfnamefont {C.-J.}\ \bibnamefont {Feng}}, \
  and\ \bibinfo {author} {\bibfnamefont {T.}~\bibnamefont {Wang}},\ }\href
  {\doibase 10.1140/epjc/s10052-024-12825-3} {\bibfield  {journal} {\bibinfo
  {journal} {Eur. Phys. J. C}\ }\textbf {\bibinfo {volume} {84}},\ \bibinfo
  {pages} {457} (\bibinfo {year} {2024})},\ \Eprint
  {http://arxiv.org/abs/2309.16944} {arXiv:2309.16944 [gr-qc]} \BibitemShut
  {NoStop}%
\bibitem [{\citenamefont {Qu}\ \emph {et~al.}(2024)\citenamefont {Qu},
  \citenamefont {Wang},\ and\ \citenamefont {Feng}}]{Qu:2023hsy}%
  \BibitemOpen
  \bibfield  {author} {\bibinfo {author} {\bibfnamefont {Z.-S.}\ \bibnamefont
  {Qu}}, \bibinfo {author} {\bibfnamefont {T.}~\bibnamefont {Wang}}, \ and\
  \bibinfo {author} {\bibfnamefont {C.-J.}\ \bibnamefont {Feng}},\ }\href
  {\doibase 10.1016/j.aop.2024.169642} {\bibfield  {journal} {\bibinfo
  {journal} {Annals Phys.}\ }\textbf {\bibinfo {volume} {464}},\ \bibinfo
  {pages} {169642} (\bibinfo {year} {2024})},\ \Eprint
  {http://arxiv.org/abs/2301.07326} {arXiv:2301.07326 [gr-qc]} \BibitemShut
  {NoStop}%
\end{thebibliography}%
	
\end{document}